\documentclass[12pt]{article}

\usepackage{epsfig}
\usepackage{amssymb}
\usepackage{multirow}
\usepackage{rotfloat}
\usepackage{amsmath,amsbsy,amsmath}
\usepackage{graphics}
\usepackage{bbm}
\usepackage{rotating}
\usepackage{lscape}
\usepackage{longtable}
\newcommand{\beq}{\begin{equation}}
\newcommand{\eeq}{\end{equation}}
\newcommand{\beqn}{\begin{eqnarray}}
\newcommand{\eeqn}{\end{eqnarray}}

\newcommand{\nn}{\nonumber}

\newcommand{\Z}{\mathbbm{Z}}

\newcommand{\be}{\begin{equation}}
\newcommand{\ee}{\end{equation}}
\newcommand{\ba}{\begin{eqnarray}}
\newcommand{\ea}{\end{eqnarray}}
\newcommand{\bdm}{\begin{displaymath}}
\newcommand{\edm}{\end{displaymath}}

\newcommand{\ie}{{\it i.e.\ }}
\newcommand{\eg}{{\it e.g.\ }}

\DeclareMathAlphabet{\mathpzc}{OT1}{pzc}{m}{it}
\def\bea{\begin{eqnarray}}
\def\eea{\end{eqnarray}}
\def\beas{\begin{eqnarray*}}
\def\eeas{\end{eqnarray*}}
\def\sla{\raise.15ex\hbox{$/$}\kern-.57em}

\def\bea{\begin{eqnarray}}
\def\eea{\end{eqnarray}}

\def\sla{\raise.15ex\hbox{$/$}\kern-.57em}
\def\ie{{\it i.e.}~}
\def\eg{{\it e.g.}~}

\def\cE{{\cal E}}

\def\cZ{{\cal Z}}

\begin{document}
%
%
\begin{titlepage}

\thispagestyle{empty}

\vspace{-2cm} 

\begin{flushright}
{\small
ROM2F/2012/07\\
}
\end{flushright}
\vspace{0.5cm}
\vskip 2cm
\begin{center}
{\Large\bf Heterotic T-folds with a small number\\of neutral moduli}
\end{center}
\vskip 2cm

\begin{center}
{\large\bf Massimo Bianchi$^a$, \,  Gianfranco Pradisi$^a$, \, \\ Cristina Timirgaziu$^b$ and Luca Tripodi$^{c}$}~\\
\vspace{0.8cm}
{\sl $^a$ Dipartimento di Fisica and Sezione INFN,\\
Universit\`a di Roma ``Tor Vergata'',\\
Via della Ricerca Scientifica 1, 00133 Roma, Italy\\
~\\
$^b$ Albert Einstein Center for Fundamental Physics\\
Institute for Theoretical Physics\\
Sidlerstrasse 5,
CH-3012 Bern,
Switzerland\\
~\\
$^c$ II. Institut f\"{u}r Theoretische Physik der Universit\"{a}t Hamburg\\
D-22761 Hamburg, Germany

}

\end{center}
\vskip 1.0cm
\begin{center}
{\large \bf Abstract}
\end{center}

We discuss non-geometric supersymmetric heterotic string models in D=4, in the framework of the free fermionic
construction. We perform a systematic scan of models with four {\it a
priori} left-right asymmetric $\Z_2$ projections and shifts. We analyze some $2^{20}$ models,
identifying 18 inequivalent classes and addressing variants generated by discrete torsions.  They do not contain geometrical or trivial neutral moduli, apart from the dilaton.  However, we show the existence of flat directions in the form of exactly marginal deformations and identify patterns of symmetry breaking where product gauge groups, realized at level one, are broken to their diagonal at higher level.  We also describe an ``inverse Gepner map'' from Heterotic to Type II models that could be used, in certain non geometric settings, to define ``effective'' topological invariants.



\vfill

\end{titlepage}

\tableofcontents
\section{Introduction}

In the past ten years the problem of moduli stabilization has attracted a lot of attention. The crucial role of internal fluxes, both for closed and open strings, has been fully appreciated \cite{flussi}. Many models with partial moduli stabilization have been proposed and non-perturbative effects, due to string or D-brane instantons, have been invoked as further means to achieve the goal \cite{reviewinstanton}. Most of the construction relies on fluxes that do not admit a full-fledged description at the world-sheet level and are only amenable to a low energy supergravity description that, among other things, requires the fluxes to be diluted. Moreover, a certain tension between chirality and moduli stabilization has been encountered \cite{TfoldsI, altri}.

Asymmetric orbifolds of tori \cite{Narain:1986qm} constitute a class of exactly solvable string models propitious to moduli stabilization. This was noted and exploited in the past \cite{Dine:1997ji} \cite{Dabholkar:1998kv} \cite{Faraggi:2005ib}, but not in a systematic manner. An extensive investigation of asymmetric orbifolds of Type II strings with very few moduli was first performed in \cite{TfoldsI} where, however,  chirality  was problematic to obtain in the unoriented descendants, because (generalized) $\Omega$ projections of left-right symmetric models based on individual left-right asymmetric twists and shifts tend to be compatible only with ``bulk'' branes.

Here, we turn our attention to the heterotic string as a more promising framework for chirality and grand unification, although the question of moduli stabilization turns out to be subtler in this setup. We build our models using the free fermionic approach developed in \cite{Kawai:1986va}  and \cite{abk}. This formulation is particularly suitable for our goal, since it allows the description of asymmetric twists and shifts in just as simple a manner as of symmetrical ones.  The free fermionic formulation of four dimensional heterotic strings has, indeed, a long tradition of semi-realistic model building \cite{nahe} \cite{ffm}. Some of the models present in the literature, made already use of asymmetric twists and shifts added to geometric orbifold projections, in the search for Standard Model like features.  They were also analyzed from the point of view of moduli fixing \cite{Faraggi:2005ib} and exhibit a reduced number of geometric moduli.  Moreover, a complete analysis of flat directions was attempted in one particular model \cite{noflat}, but, due to the extension of the moduli space, it was difficult to reach a definite conclusion, even using computer tools.

In this paper, we investigate in a systematic way the outcome of combining the virtues of semi-realistic NAHE models \cite{nahe} with asymmetric twists and shifts, in a simple setup. We scan a large class of asymmetric heterotic string models, employing  as a starting point \textit{purely chiral} twist and shifts that give rise to NAHE-like sectors. Geometrical moduli are \textit{always} fixed in this framework and no obvious neutral scalar moduli appear. However, this does not guarantee the lack of flat directions. Combining an algorithm to scan the possible models and analytic considerations, we find 18 classes of supersymmetric models, each equipped with $2^6$ discrete torsion variants. Some of them exhibit the GUT $SO(10)$ gauge group and chirality. Semi-realistic vacua would require more involved model building, where extra projections further break the gauge group and reduce the number of Standard Model generations. We defer such an analysis to the future and focus on models with four {\it a priori} left-right asymmetric $\Z_2$ projections and no Ising fermions.\footnote{For an analysis involving geometrical projections, see \cite{Assel:2010wj}.}
The exhaustive examination of flat directions in a heterotic string model is also a very complex problem. We only address a qualitative analysis, that basically demonstrates how the existence of exactly marginal deformations cannot be avoided in this class of models.

The paper is organized as follows. In Section 2 we review the basics of the free fermionic construction of heterotic string models. In Section 3 we analyze the possibility of building heterotic asymmetric models with standard embedding, via the so called Gepner map, starting from the Type II models obtained in \cite{TfoldsI}. For later purposes we also define an ``inverse'' Gepner map that allows to relate heterotic models with gauge symmetry $G\supset SO(10)\times E_8$ to Type II models with at least $N=1$ supersymmetry. Section 4 details the construction of our models and the results. A comparison with the Type II models obtained in \cite{TfoldsI} is included together with an analysis of the ``effective'' topology of the heterotic vacuum gauge bundle ${\cal E}$. Finally, Section 5 offers a discussion of the moduli in free fermionic heterotic models of the NAHE kind and of the heterotic analogue of brane recombination, whereby product groups are broken to their diagonal, with current algebras realized at higher level. The paper also contains several appendices. Appendices A and B depict an analytical derivation of our models, while in appendix C we discuss the effect of turning on discrete torsion in one particular model. Appendices D and E contain the technical details relevant for the writing of the modular invariant partition functions, as well as for the algorithm used to scan our $2^{20}$  models via a Mathematica program.

\section{Free fermionic construction}\label{Sec:freefermions}

In the free fermionic construction \cite{Kawai:1986va} \cite{abk} of four-dimensional heterotic string models all the internal degrees of freedom are represented in terms of fermions. The resulting {\it rational} Conformal Field Theories have the virtue of being relatively simple, while loosing, in the general case, an immediate geometric interpretation.  The left-moving degrees of freedom correspond at least to an $N=1$ world-sheet supersymmetry.  They result into
$18$ internal (real) fermions $\chi^i$, $y^i$ and $\omega^i$, $i=1,\ldots, 6$, besides the two uncompactified (light-cone) coordinates $\partial X_L^{\mu}$ and their superpartners $\psi^{\mu}$.  The $\chi^i$ are the fermionic coordinates along the compact directions, while $y^i$ and $\omega^i$ fermionize the compact space (chiral) bosonic coordinates according to
\be \partial X_L^i = y^i \omega^i,  \quad\quad i=1,\ldots  , 6 \ , \label{bosoniz1}\ee
in such a way that the world-sheet supercurrent can be non-linearly realized as
 \be G \, = \, \sum_{\mu=7,8} \, \psi^\mu \, \partial X_{L \, \mu} \, + \, \sum_{i=1}^6 \,  \chi^i y^i \omega^i. \quad \label{susy}\ee
The right-moving degrees of freedom include the uncompactified (light cone) coordinates $\bar{\partial} \tilde{X}_R^{\mu}$
together with $44$ internal real fermions.  We label them as follows: $\tilde{y}^i$ and $\tilde{\omega}^i$, in analogy with their left-moving companions, are the fermionization of the internal compact space (chiral) bosonic coordinates
\be \bar{\partial} \tilde{X}_R^i = \tilde{y}^i \, \tilde{ \omega}^i,  \quad\quad i=1,\ldots  , 6\ . \label{bosoniz2}\ee
It is customary to split the remaining 32 real fermions as ($\tilde{\chi}^{1...6}, \, \tilde{\psi}^{1...10},\, \tilde{\phi}^{1...16}$), where the $\tilde{\chi}$'s generate a third internal $SO(6)$ that proves to be useful in order to identify the ``standard embedding'', while the
$\tilde{\psi}^{1...10}$ and the $\tilde{\phi}^{1...16}$ are related to the ``visible'' $SO(10)$ and the ``hidden'' $E_8$ in the semi-realistic four-dimensional model building.

To construct a theory, we need to specify a basis of fermion sets, $\{b_\alpha\}$.  Each set contains the fermions that are grouped together with identical spin structure in the corresponding projection.  In the real fermion case, that will be our main focus, one can only have $\Z_2$-valued phases.  As shown in the original papers \cite{abk}, the sets form a group with identity (the empty set) under the symmetric difference. With a certain abuse of notation, we will indicate this operation with the symbol ``+'' \ie
\be
b_1+b_2 = \{f^a \} \quad \iff \quad  \left[ f^a \in  b_1\cup b_2\right] \wedge \left[ f^a \notin b_1\cap b_2\right] \ .
\label{symmdiff}
\ee
By the same token, we will sometimes refer to the sets as ``vectors''.

Not all the possible choices of basis are of course compatible with a consistent superstring model.  Indeed, modular invariance translates into the conditions
\bea
   n(b_\alpha)&=&0~{\rm mod }~8 \ ;\nn\\
    n(b_\alpha \cap b_\beta)&=&0~{\rm mod }~4 \ ;\nn\\
    n(b_\alpha \cap b_\beta \cap b_\gamma)&=&0~{\rm mod }~2 \ ;\nn\\
       n(b_\alpha \cap b_\beta\cap b_\gamma \cap b_\sigma)&=&0~{\rm mod }~2 \ ,
\label{consistency}
\eea
where $n(b)$ denotes the difference between the number of left- and right- moving fermions in the set $b$.
Additionally, preservation of the holomorphic world-sheet supercurrent is also a necessary condition, that translates into
\be
  \forall i \, \quad\quad   \# ~\chi^i - \# ~y^i -\# ~ \omega^i=0~{\rm mod }~2 \ ,
  \label{condsusy}
\ee
where $ \#$ indicates the number of the corresponding fermions.
To the previous constraints, one has to add the correct relation between spin and statistics, derived from the factorization of higher loop amplitudes \cite{abk,MBASrelmodgr}.
Having $m$ basis vectors results into a space of $2^m$ sectors. The states in each sector have to be submitted to the projections related to the initial sets.  Discrete torsions play also an important role in building the final spectrum.

The simplest models, those with a basis containing few sets, correspond to heterotic strings at enhanced symmetry points, where the background metric and antisymmetric tensor take specific values. For instance, the $E_8 \times E_8$ heterotic string compactified on the maximal torus of $SO(12)$ is obtained from the sets
 \bea   F &=& \{ \psi^\mu\, \chi^{1\ldots 6} \, y^{1\ldots 6} \,
 \omega^{1\ldots 6}  | \, \tilde{y}^{1\ldots 6}\,\tilde{ \omega}^{1\ldots 6}\, \tilde\chi^{1\ldots 6}\, \tilde\psi^{1...10}\, \tilde\phi^{1...16}   \} \ , \nn\\
S & =& \{\psi^\mu\, \chi^{1\ldots 6}  \}\ , \nn\\
{E} &=&\{ \tilde\phi^{1...16} \}\ , \nn\\
{G} &=&\{\tilde\chi^{1\ldots 6}\, \tilde\psi^{1...10}\}\ .
\label{basiszero}
\eea
The set $S$ is responsible for the usual GSO projection. The basis $\{F, S\}$ generates the partition function of the Narain generalized toroidal compactification of the heterotic string \cite{Narain:1985jj}.  The corresponding one-loop partition function, omitting the integration over the moduli space, can be written in the form\footnote{We follow the notation and conventions of \cite{AngSagn}.}
\bea {\cal T} = {1\over \eta^2 \bar \eta^2} \left(V_8-S_8\right)\left(O_{12}\bar O_{44}+V_{12}\bar V_{44}+S_{12}\bar S_{44}+C_{12}\bar C_{44}\right)\ .\label{FS}\eea
The four dimensional massless spectrum is the one of an $N=4$ supergravity coupled to an $N=4$ Super-Yang-Mills theory with an $SO(44)$ gauge group.  It should be noticed that the geometric massless scalars in this model parameterize the coset
$SO(6,22)/[SO(6)\times SO(22)]$  and are in one-to-one correspondence with the $6\times 22$ components of the background metric, antisymmetric tensor and Wilson lines \cite{Narain:1985jj}.  They correspond to the scalars in the Cartan subalgebra of $SO(44)$.

Adding ${E}$ operates the separation of the hidden gauge group and its enhancement from
$SO(16)$ to $E_8$\footnote{For a particular choice of discrete torsions.}.  The partition function
\bea {\cal T} = {1\over \eta^2 \bar \eta^2} \left(V_8-S_8\right)\left(O_{12}\bar O_{28}+V_{12}\bar V_{28}+S_{12}\bar S_{28}+C_{12}\bar C_{28}\right)\left(\bar O_{16} + \bar S_{16}\right)\ \label{starting point}\eea
exhibits neatly the $N=4$ model with an $E_8\times SO(28)$ enhanced gauge group.  This (generalized) toroidal compactification will be our starting point to perform the asymmetric orbifold projections. Notice that $\bar O_{16} + \bar S_{16} = \bar{\cE}_8$ is the character of $E_8$ at level one.
Finally, let us remark that by adding ${G}$, the compact space degrees of freedom are separated from those generating the visible $E_8$ producing the announced compactification on the $SO(12)$ maximal torus
        \bea {\cal T} = {1\over \eta^2 \bar \eta^2} \left(V_8-S_8\right)\left( |O_{12}|^2 + |V_{12}|^2 + |S_{12}|^2+|C_{12}|^2\right)\left(\bar O_{16} + \bar S_{16}\right)\left(\bar O_{16} + \bar S_{16}\right)\ .\eea

\section{(Inverse) Gepner map and (non) standard embedding}

In \cite{TfoldsI} a combination of chiral twists and shifts has been used in Type IIB  asymmetric orbifolds and corresponding unoriented descendants to produce (super)string vacua with a small number of moduli, in the framework of the free fermionic construction.  The idea was to use shifts to lift in mass twisted moduli and chiral twists to unpair the untwisted ones.  As a result of a scan
over a huge number of possible vacua, many examples of ``effective Calabi-Yau compactifications'' with small Hodge numbers were found, including an $N=2$ supersymmetric ``self-mirror'' model with $(h_{11}=1, h_{12}=1)$. An $N=1$ Type I theory without open strings was also described, whose $\Omega$ projection kept the vector boson in the vector multiplet of the parent type II theory. The massless spectrum contains just the $N=1$ supergravity multiplet, the dilaton chiral multiplet and an additional chiral multiplet: it is, as far as we know, the minimal spectrum one can get for a superstring vacuum. Minimal models with $N=2$ and $N=3$ supersymmetry have been derived in \cite{DJK,FK,KLS,Candav,tretremodels}.

As well known, the interesting semi-realistic vacua in the heterotic description are exactly those with both Hodge numbers small, see \eg \cite{Candav, tretremodels, candelas} and references therein.  The natural attempt to get interesting models with few moduli is thus to apply to the heterotic strings a construction similar to the one in \cite{TfoldsI}.  There is a procedure, usually called the ``Gepner map'', that allows one to get a consistent heterotic compactification with ``standard embedding'' of the spin connection in the gauge group, starting from a consistent Type II compactification \cite{Gepner,SY}\footnote{We thank A.N. Schellekens for having called to our attention the fact that the ``Gepner map'' was applied in its full-fledged form in \cite{SY}.}.
The easiest way to describe the Gepner map in our notation is using characters or super-characters.
In terms of characters, the map leaves the left-movers untouched and acts on the right-movers according to
\be
V_2^{st} \rightarrow O_{10}\times E_8\ ; \quad O_2^{st} \rightarrow V_{10}\times E_8 \ ;
\quad -S_2^{st} \rightarrow +S_{10}\times E_8 \ ; \quad -C_2^{st} \rightarrow +C_{10}\times E_8 \ ,
\ee
for the $E_8\times E_8$ heterotic string, or
\be
V_2^{st} \rightarrow O_{26} \ ; \quad O_2^{st} \rightarrow V_{26} \ ;
\quad -S_2^{st} \rightarrow +S_{26} \ ;  \quad -C_2^{st} \rightarrow +C_{26} \ ,
\ee
for the $Spin(32)/\Z_2$ heterotic string. The sign change for both spinors is due to spin and statistics, and the $O\leftrightarrow V$ flip is consequently required by modular invariance.

In terms of supercharacters, up to the $E_8$ factor,
\begin{equation}
v= V_2 \xi_0 + O_2 \xi_3 - S_2 \xi_{3/2} - C_2 \xi_{-3/2} \rightarrow
a= O_{10} \xi_0 + V_{10} \xi_3 + S_{10} \xi_{3/2} + C_{10} \xi_{-3/2} \ , \nonumber \end{equation}
\begin{equation}
\phi= V_2 \xi_{-2} + O_2 \xi_{+1} - S_2 \xi_{-1/2} - C_2 \xi_{+5/2} \rightarrow
t = O_{10} \xi_{-2} + V_{10} \xi_{+1} + S_{10} \xi_{-1/2} + C_{10} \xi_{+5/2} \ , \nonumber \end{equation}
\begin{equation}
\phi^c= V_2 \xi_{+2} + O_2 \xi_{-1} - S_2 \xi_{-5/2} - C_2 \xi_{+1/2} \rightarrow
t^c = O_{10} \xi_{+2} + V_{10} \xi_{-1} + S_{10} \xi_{-5/2} + C_{10} \xi_{+1/2} \ , \end{equation}
where $\xi_q$ are the characters of the $N=2$ minimal model at $c=1$, equivalent to a compactified boson at radius $r=\sqrt{3}$.
It is easy to see that $SO(10)\times E_8$ gets enhanced to $E_6\times E_8$, since
\be
a=\chi_{1}^{E_6} \ , \qquad t=\chi_{27}^{E_6}\ , \qquad t^c=\chi_{27^*}^{E_6} \ .
\ee
In particular, at the massless level ${\bf 78} \rightarrow {\bf 45}_0 + {\bf 1}_0 + {\bf 16}_{+3/2} + {\bf 16}^*_{-3/2} $
while ${\bf 27} \rightarrow {\bf 1}_{-2} + {\bf 10}_{+1} + {\bf 16}_{-1/2} $. For $Spin(32)/\Z_2$ there is no enhancement of $SO(26)\times U(1)$.

It is important to stress that in geometric contexts the ``Gepner map'' always produces
heterotic models with standard embedding and chiral asymmetry $N_{\bf 27} - N_{\bf 27^*} = h_{11} - h_{21}$
or $N_{{\bf 26}_{+1}} - N_{{\bf 26}_{-1}} = h_{11} - h_{21}$.
 As a result, the most interesting models of \cite{TfoldsI} with identical Hodge numbers tend to be non chiral, unless the presence of disjoint orbits under modular transformations allows to apply different Gepner maps in different sectors and get chiral models even if $h_{11} = h_{21}$, as we will see later on.  Moreover,
the neutral moduli are as in Type II plus a number of (charged) singlets and deformations of the gauge bundle corresponding to $H^1(EndT)$.

Barring the effect of discrete torsion, it is more
convenient to try and use chiral twists and shifts in frameworks that look phenomenologically more promising.  The one we are going to exploit in the next sections is inspired by the semi-realistic class of models known as NAHE \cite{nahe}. As we will see, these do not admit an immediate geometric interpretation. However, as in Type II models, one can try to {\it define} effective topological numbers associated to the chiral massless spectrum. For instance, one could define an ``inverse'' Gepner map. It would work whenever the gauge
group contains a factor $SO(10)\times E_8$, as in some of the 18 classes of interesting models that we analyze. Indeed in this case one could always map heterotic characters
into Type II characters according to \footnote{Depending on the choice of chirality from $SO(10)_{het}$ to $SO(2)_{st}$, one gets Type IIB or Type IIA models.}
\be
O_{10}\times E_8 \rightarrow V_2^{s-t}\ , \quad
V_{10}\times E_8 \rightarrow O_2^{s-t} \ , \quad
S_{10}\times E_8 \rightarrow -S_2^{s-t} \ , \quad
C_{10}\times E_8 \rightarrow -C_2^{s-t} \ . \label{invegepmap} \ee
The resulting Type II model is perfectly consistent and modular invariant though {\it a priori} non geometric.
By construction, it enjoys at least ${N}=1$ supersymmetry, \ie ${N}_L=1$ and ${N}_R = 0$. In some cases,  enhancement to ${N}=2$ supersymmetry, namely ${N}_L=1$ and ${N}_R = 1$, can take place that  allows one to define effective Hodge numbers as in \cite{TfoldsI}. This happens when in different sectors charged matter appears in representations of different $E_6$'s. One can then go back to the heterotic model and define ``effective'' topological numbers of the vacuum gauge bundle ${\cal E}$ that should be interpreted as a non-geometric version of the (non) standard embedding. Not all is lost in the generic case since the gauge group in the Type II model is abelian and one can count $N=1$ vector and neutral chiral multiplets. The latter could be identified as neutral moduli.

\section{Models with four $\Z_2$ L-R asymmetric projections}

In this section, we present our systematic scan for heterotic models with four {\it a
priori} left-right asymmetric $\Z_2$ projections that eliminate neutral moduli. Starting with NAHE-inspired models, we analyze $2^{20}$ models, identify 18 inequivalent classes of non-geometric models (``T-folds'' in certain cases) and then address variants generated by discrete torsions.

\subsection{NAHE-inspired models}

The original NAHE class of models \cite{nahe} is a $\Z_2\times \Z_2$ orbifold of the original $SO(12)_L \times SO(28)_R \times E_{8R}$ described in eq. (\ref{starting point}).  The projection breaks the symmetry to
$SO(4)^3_L \times SO(10)_R \times SO(6)^3_R \times E_{8R}$.  In the fermionic construction, the orbifold can be realized using the basis sets $\{F, S, {E}\}$ of eq. (\ref{basiszero}), together with the two additional ones
\bea
&& b_1 = \{\psi^\mu\, \chi^{1,2}\, y^{3..6} | \, \tilde{y}^{3..6}\, \tilde{\chi}^{1,2} \,\tilde{\psi}^{1..10}    \}\ , \nn \\
&& b_2 = \{\psi^\mu\, \chi^{3,4}\, y^{1,2}  \omega^{5,6} |  \,\tilde{y}^{1,2} \,\tilde{ \omega}^{5,6}  \, \tilde{\chi}^{3,4} \, \tilde{\psi}^{1..10} \}\ , \eea
or, equivalently, using the sets $\{F, S, b_1, b_2, b_3\}$, with
 \be b_3=F+b_1+b_2+{E}= \{\psi^\mu\, \chi^{5,6}\, \omega^{1..4} | \, \tilde{\omega}^{1..4}\, \tilde{\chi}^{5,6} \,\tilde{\psi}^{1..10}    \} \ . \ee
Additional sets can be added to the basis in order to produce semi-realistic models whose spectra are close to the one of the Standard Model or some of its GUT's or supersymmetric extensions.  In the original paper, for instance, three more sets provide the breaking of $SO(6)^3$ to $U(1)^3$ and of $SO(10)$ to $SU(5)\times U(1)$, with a resulting ``visible'' gauge group $SU(5)\times U(1) \times U(1)^3$, three generations of quarks and leptons and additional exotic matter.  These models are in the class of ``flipped'' $SU(5)$ GUT's.

 At the level of the NAHE set, the geometrical moduli that survive the orbifold projection appear to be charged, but an analysis of the tree level super-potential reveals the presence of flat directions.  We will come back to this issue in the last Section.


\subsection{Our approach}\label{subs:ourapproach}
As noted in \cite{Faraggi:1990ac}, when looking for chiral spinorial representations of $SO(10)$, which do not carry charges under the hidden gauge group, one is led naturally to the NAHE set. That is to say, the vectors
$b_1$, $b_2$ and $b_3=1+b_1+b_2+{E}$ have the correct form to give rise to a certain number of copies of the $16$ representation  of $SO(10)$. Chirality, in particular, is insured thanks to the fact that the $b_i$ vectors only share the fermions $ \psi^\mu $ and $\tilde{\psi}^{1..10}$. If this were not the case, the projection by $b_1$ in the sector twisted by $b_2$, for instance, would lead to an equal number of $16$ and $\overline{16}$ of SO(10).

Hence, for a model to be semi-realistic, one should aim at having such sectors in the Hilbert space. Aside from building models based on the NAHE set, one can consider the option of obtaining NAHE-type vectors as combinations of the initial set of basis vectors. In the following we exploit this second option.

We consider models based on the sets $\{ F,S, {E} \}$ together with four additional sets. As explained before, ${E}=\{\tilde{\phi}^{1...16}\}$ is equivalent to a Wilson line performing the separation of the hidden gauge group. Additionally, the hidden gauge group is enhanced from $SO(16)$ to $E_8$ by massless gauge bosons in the $128$ (spinorial) representation of $SO(16)$,  generated in sector ${E}$ \footnote{This is dependent on the choice of the discrete torsions. In the following we set the relevant discrete torsions to the values that allow the mentioned enhancement.}. We want to combine the virtues of the NAHE-type sets with the advantages of asymmetric orbifolds and the possibility of adding shifts that lift in mass the twisted moduli.  The four additional sets assume the form
\bea &&
b_1 =(b_{1L},b_{1R})= I_{3456}\, \sigma^{i_1 i_2 \ldots }\,\bar \sigma^{k_1 k_2 \ldots } = \{ (\chi \, \omega)^{3456}  \, (y\, \omega)^{i_1 i_2 \ldots }  | (\tilde y\, \tilde \omega)^{k_1 k_2 \ldots }  \} \ , \nn\\
&&b_2 =  (b_{2L},b_{2R})=I_{1256}\, \sigma^{j_1 j_2 \ldots }\,\bar \sigma^{l_1 l_2 \ldots } = \{(\chi\,  \omega)^{1256}  \, (y\, \omega)^{j_1 j_2 \ldots }  | (\tilde y\, \tilde \omega)^{ l_1 l_2 \ldots }   \} \ , \nn\\
&&b_3 =(b_{3L},b_{3R})= \bar I_{3456}\, \sigma^{k'_1 k'_2 \ldots }\,\bar \sigma^{i'_1  i'_2 \ldots }  = \{ ( y\,  \omega)^{k'_1 k'_2 \ldots }  | (\tilde\chi\, \tilde  \omega)^{3456}
  (\tilde y\, \tilde \omega)^{i'_1 i'_2 \ldots }     \} \ , \nn\\
&&b_4  =(b_{4L},b_{4R})= \bar I_{1256}\, \sigma^{l'_1 l'_2 \ldots }\,\bar \sigma^{ j'_1  j'_2 \ldots } = \{ ( y\,  \omega)^{ l'_1 l'_2 \ldots } | (\tilde\chi\, \tilde  \omega)^{1256}
(\tilde y\, \tilde \omega)^{j'_1 j'_2 \ldots }  \} \  , \label{ansatz1}\eea
where $I_i$ and $\sigma_i$ correspond to the reflections
\bea
I_i = \{\chi^i \, \omega^i \}  : && \chi^i \to -\chi^i \ , \quad \quad   \omega^i \to - \omega^i  \ ;  \nn\\
\sigma_i=\{y^i \,\omega^i \}: &&   y^i \to -y^i \ , \quad \quad  \omega^i \to -\omega^i  \ ;
\eea
and correspondingly for $\bar I_i$ and  $\bar \sigma_i$. In view of equations (\ref{bosoniz1}), (\ref{bosoniz2})
one can observe that $I_i$ acts as a $\Z_{2L}$ {\em chiral} reflection of the $i^{\rm th}$ left-moving internal bosonic and fermionic coordinates (and equivalently for $\bar I_i$)
\bea
 I_i: && X_L^i   \rightarrow  - X^i_L \ ,
 \quad \quad X_R^i
\rightarrow  X^i_R \ ,  \eea
while $\sigma_{i}$ stands for a left moving {\em chiral} shift along the $i^{\rm th}$ direction (and likewise for $\bar \sigma_i$)
 \be
 \sigma_{i}: X^i_L \to X^i_L+\delta \ , \qquad  X^i_R \to X^i_R \ ,
 \ee
with $2\delta$ a chiral lattice vector. Chiral reflections are T-duality transformations. Thus one can dub the models we find ``Heterotic T-folds'', very much as the models in \cite{TfoldsI}, deserve to be dubbed Type II T-folds.

Sets (\ref{ansatz1}) are very similar to the ones considered in \cite{TfoldsI}, with the notable difference that in the type IIB case the sets $b_3$ and $b_4$ were exactly the mirrors of $b_1$ and $b_2$. The choice of identical actions on the left and right movers was justified there by the prospect of performing an unoriented projection and including D-branes and open strings for type I models.  In the present work, we can relax that condition and allow for  more general shifts.  We remark that, at this first stage, we limit ourselves to sets for which the fermions are arranged into pairs with identical spin structure.  In this way we avoid the so-called Ising fermions, that provide interesting models, but reduce the rank of the gauge group giving rise to additional neutral moduli. The scan is realized analyzing all the possible combinations of the indices $(i,j,k,l,i',j',k',l')$ compatible with the constraints illustrated in Section 2 and the pairing of fermions.
NAHE-type vectors are potentially obtained from combinations of the basis sets such as\footnote{Sectors of the form $F+b_i+E$, altougth not of NAHE type, can also contribute spinorials of $SO(10)$ in models with 6-shifts.}
  \bea &&F+b_1+b_{3,4}+{E} \ , \nonumber \\
  &&F+b_2+b_{3,4}+{E}\ , \nonumber \\
   &&F+b_i+b_j+ b_k+ {E}\ , \quad i\neq j \neq k \ ,\nonumber \\
   &&F+b_1+b_2+b_3+b_4+{E}\ .
  \eea
Given the sets, one has to build the modular invariant one-loop partition function and extract the massless spectra.  In Appendix \ref{characters} we report the supersymmetric characters $\tau_{ij}$ (in terms of the characters of $SO(2)^{\otimes 4}$) used to deal with the left-moving GSO projection related to the chiral $\Z_2\times\Z_2$ twists.  They realize the desired orbifold decomposition of the $V_8 - S_8$ term \cite{MBthesis} \cite{MBAS}.  The rest is obtained using the twists on the theta functions
\be
\theta_1 \rightarrow -\theta_2\ ; \quad \theta_2 \rightarrow \theta_1\ ;\quad \theta_3 \rightarrow \theta_4\ ;\quad \theta_4 \rightarrow \theta_3\ .
\label{twistontheta}
\ee
This way of building the amplitudes has the advantage of incorporating automatically the spin-statistic connection. The complete expressions, however, are clever combinations of the different contributions.  The modular invariant partition function, limited to the fermion contributions, has an expression of the form
\be
\cZ \ = \ \frac{1}{16} \ \sum_{\alpha, \beta} \ C_{\alpha \beta} \ \rho_{\sigma_{\alpha} \sigma_{\beta}} \ \Lambda_{\alpha \beta}\ ,
\ee
where $\alpha$ and $\beta$ run over all the sets, $\rho_{\sigma_{\alpha} \sigma_{\beta}}$ are the suitable combinations of $\tau$'s corresponding to the twist in the $(\alpha, \beta)$ sector and $\Lambda_{\alpha \beta}$ are the amplitudes related to the remaining $56$ fermions obtained with the rules (\ref{twistontheta}) and $S$ and $T$ modular transformations.  The coefficients $C_{\alpha \beta}$ are signs, to be chosen in such a way that $\cZ$ be modular invariant.  The orbifold group is built out of $16$ sets, corresponding to as many sectors, for a total of $256$ amplitudes.  They are organized into $36$ modular orbit: the untwisted orbit of length $46$ ($16$ untwisted plus $2 \times 15$ twisted amplitudes) and $35$ additional disconnected orbits of length $6$.  Of course, not all of them are independent since they have to respect certain quadratic constraints due to the fact that in each sector one has to get a projection operator \cite{abk}.  As a result, the only independent coefficients are $C_{b_i,b_j}, i>j$, where ${b_i,b_j}$ are elements of the basis.  In our case, the basis is of $5$ elements, with $10$ independent coefficients (besides the initial $C_{00}=1$ related to the empty set or to the identity). Four of them, however, fix just the untwisted projection and will be set to $1$. As a consequence, each modular invariant has $2^{6}$ discrete torsion variants.  Moreover, not all the independent models give rise to distinct physical vacua.  Many of them are indeed equivalent, resulting just in a reshuffling of the ordering of the internal fermions. In the next section we will address this issue in detail.

\subsection{Our models}

In our case, the $\Z_2 \times \Z_2$ {\em chiral} orbifold actions combined with the shifts project out all the moduli in the
Cartan subalgebra.  Moreover, since we do not have rank reduction of the gauge group, we can exclude flat directions with neutral moduli, but we will see that there remain flat directions along bi-fundamental fields that break product groups to their diagonals.

Depending on the details of a model and on the discrete torsions, some of the NAHE-like sectors might contribute massless states. Unfortunately, the relation between the chirality of the states from these sectors and the discrete torsions is more involved than in the case of the NAHE set. However, as explained in appendix \ref{DTsection}, it is still interesting to obtain models where  the NAHE-like sectors overlap only in the fermions $\{ \psi^\mu ~|~ \tilde \psi^{1...10}  \}$.
Other sectors producing massless states might be present in the theory as well. 

At the end of the analysis, we find $256$ models falling into $18$ inequivalent classes of heterotic-string vacua without neutral moduli, barring the dilaton.  In appendices \ref{sets} and \ref{ineq} we explain in detail how to derive them from our basis vectors (\ref{ansatz1}).  In the following we sketch the main points.

Since the twists are already fixed, we need to determine the inequivalent shifts. Let us first discuss the shifts that accompany the twists, for instance the shifts in the left-moving component of $b_{1,2}$.  Certain shifts are redundant when combined with certain twists. This restricts the independent choices of $b_{1L}$ and $b_{2L}$. The right-moving shifts in $b_{1,2}$ are also partially\footnote{The number of shifted fermions is restricted, but not which ones.} restricted by the modular invariance of $b_1$ and $b_2$.
We end up with 8 choices for $b_1$ and 16 choices for $b_2$. To implement the modular invariance constraint, $b_1\cdot b_2=0~ \rm{mod} ~4$, one needs to look at the left-movers in $b_1$ and $b_2$, because the right-movers always contribute a multiple of four. It turns out that this modular invariance condition cuts by half the number of possibilities. The analysis can be repeated for $b_3$ and $b_4$, yielding
$2^{12}$ different models. As we expect, the other modular invariant constraints reduce this number to the $2^8=256$ distinct models found by implementing the algorithm described in Appendix E on a Mathematica program. Hence, a reduction by a factor of  $2^4$, corresponding to the four modular invariance conditions  $b_{1, 2} \cdot b_{3,4}$\footnote{Notice that some modular invariance conditions, such as $n(b_i \cap b_j \cap b_k)=0 ~\rm{mod}~ 2$, are automatically satisfied, because Ising fermions are excluded.}, is at work.

Next we investigate the inequivalent values for the shifts in the right-moving components of $b_{1,2}$ and left-moving components of $b_{3,4}$. To this end, we remark that a given set $\{b_1, b_2, b_3,b_4\}$  carries certain symmetries that transform it into an equivalent set of the form (\ref{ansatz1})\footnote{For instance switching the right-moving components of the shifts between $b_1$ and $b_2$.}.
These transformations, detailed in equation (\ref{sym}) in appendix \ref{sets}, form a group with 36 elements (with the identity). Hence, each class of equivalent models should contain a priori 36 elements. However, in most cases, some of the previous symmetries are trivial due to the particular form of the basis vectors (for instance lack of shifts) and, as a result of this, some classes contain fewer elements. Combining the various possible values of the shifts one can discriminate all possible inequivalent models.

The resulting models are gathered in Table  \ref{tabmodels}.
The table details, for each model, the basis sets, the gauge group and the amount of supersymmetry. For those cases in which the enhancement of the gauge group and/or of supersymmetry from twisted sectors is possible, the table displays the gauge group and the amount of supersymmetry obtained from the untwisted sector, as well as the enhanced gauge group and supersymmetry obtained without discrete torsion. 

In models with an $SO(10)$ gauge group, with the exception of model 3,  the enhancement of the gauge group can be prevented with an appropriate choice of discrete torsion. In the case of model 3, there are three sectors contributing gauge bosons: $b_3$, $b_4$ and $b_3+b_4$. The extra gauge bosons charged under $SO(10)$, coming from sector $b_3$, can be projected out choosing $C(b_2,b_3)=-1$. Those from sector $b_4$ can be removed with $C(b_2,b_4)=-1$, while sector $b_3+b_4$ yields no gauge bosons charged under $SO(10)$ for $C(b_2,b_3) \cdot C(b_2,b_4)=-1$. Hence, for any choice of discrete torsion, at least one twisted sector leads to an enhancement of the $SO(10)$ gauge group. 

The enhancement of supersymmetry can also be prevented for certain values of the discrete torsion and it is interesting to note that keeping an $SO(10)$ gauge group is compatible with keeping $N=1$ supersymmetry in the relevant models. Indeed, in models 4, 5, 11 and 12 the enhancement of the $SO(10)$ gauge group can be prevented by imposing $C(b_2,b_4)=-1$, while avoiding enhancement of supersymmetry in models 3 to 10 requires $C(b_2,b_3)=-1$ or $C(b_2,b_4)=-1$. In models 1 and 2 suppressing the enhancement of supersymmetry involves also the phases $C(b_1,b_3)$ and $C(b_1,b_4)$.


Let us analyze more in detail the effects of turning on discrete torsion and make a few remarks about the connection with phenomenology of our models. We have found 9 models with $N=1$ supersymmetry that feature an $SO(10)$ gauge group, models 4, 5, 10, 11, 12, 13, 14, 17 and 18. Among them, models 4, 11 and 17
 can be easily identified as being non-chiral, whatever the choice of the discrete torsion is. For instance, the sets of models 11 and 17 do not contain the left moving fermions $y_{56}$. Because of this the $SO(10)$ fermions $\tilde\psi^{1...10}$ are always paired with $y_{56}$ and, as a result, always lead to $16+\overline{16}$ representations. The case of model 4 is more involved, but it can be checked that in each twisted sector contributing spinorial representations of $SO(10)$ there is an excess of variables to be fixed with respect to the conditions imposed.

 Other models, for instance 13 and 14, are chiral for all choices of discrete torsion and the net chirality is independent on it. Models with no shifts in the right part of $b_1$ and $b_2$ can potentially contain adjoint scalars in the twisted sectors, as well as states charged both under the visible and the hidden gauge groups.  The presence of such states and other exotic fields can sometimes be controlled by the choice of discrete torsion and one can speculate that by adding extra basis vectors to a given model one can render it more "realistic". Since the exhaustive analysis of the effects on the spectrum of the discrete torsion is rather lengthy and very model-dependent, we investigate these effects in full detail only for model 13. It  has the nice feature of exhibiting only three twisted sectors that contribute chiral states, each one with four $16$'s of $SO(10)$, while the other  twisted sectors contributing spinorials always contain an equal number of $16$'s and $\overline{16}$'s.  Details are reported in appendix \ref{DTsection}.

{\footnotesize

\begin{longtable}{|c|c|c|c|c|}
\caption{List of the 18 classes of independent models.}\label{tabmodels}\\

\hline
\textit{M}  & \textit{Sets} & \textit{ Action} & \textit{  Gauge group Untw/no DT} & $N$ \\
\hline
\endfirsthead

\multicolumn{5}{c}{\bfseries \tablename \ \thetable{} -- continued from previous page}\\
\hline
\textit{M}  & \textit{Sets} & \textit{Action} & \textit{Gauge group Untw/no DT} & $N$ \\
\hline
\endhead

\hline\multicolumn{5}{|r|}{Continued on next page}\\
\hline
\endfoot

\hline
\hline
\endlastfoot

\textbf{1} &  $b_1=  \{ \chi_{3456}\   \omega_{3456}\  \ \|\   \}$ & $I_{3456}$ &$SO(4)^3\otimes SO(16)\otimes \ E_8$& $2$\\
\textbf{}  & $b_2=  \{ \chi_{1256}\ \omega_{1256}\  \ \|\ \}$ & $I_{1256}$ &$~$&~\\
\cline{4-5}
\textbf{}  & $b_3=  \{  \|\   \tilde{\chi}_{3456}\   \tilde{\omega}_{3456}\ \}$ & $\bar{I}_{3456}$ & $SO(28)\otimes E_8$ & 4\\
\textbf{}  & $b_4=  \{ \ \|\   \tilde{\chi}_{1256}\  \tilde{\omega}_{1256}\  \}$ & $\bar{I}_{1256}$ & ~ &~\\
\hline
\hline
\textbf{2} & $b_1=  \{ \chi_{3456} \ \omega_{3456}\  \ \|\   \}$ & $I_{3456}$ &$SO(4)^2\otimes SO(8)\otimes \ SO(12)$& $2$\\
\textbf{} & $b_2=  \{ \chi_{1256}\   \omega_{1256}\  \ \|\   \}$ & $I_{1256}$ &$\otimes \ E_8$&~\\
\cline{4-5}
\textbf{}  & $b_3=  \{ y_{123456}\  \ \omega_{123456}\  \ \|\   \tilde{\chi}_{3456}\  \ \tilde{y}_{12}\  \ \tilde{\omega}_{123456}\ \}$ & $\sigma_{123456}\bar{I}_{3456}\bar{\sigma}_{12}$ &$SO(12) \otimes SO(16) \otimes E_8$ & 4\\
\textbf{}  & $b_4=  \{  \ \|\   \tilde{\chi}_{1256}\  \ \tilde{y}_{56}\  \ \tilde{\omega}_{12}\ \}$ & $\bar{I}_{1256}\bar{\sigma}_{56}$ & ~&~\\
\hline
\hline
\textbf{3} & $b_1=  \{ \chi_{3456}\  \ y_{12}\  \ \omega_{123456}\  \ \|\   \tilde{y}_{123456}\  \ \tilde{\omega}_{123456}\ \}$ & $I_{3456}\sigma_{12}\bar{\sigma}_{123456}$ &$SO(2)^6\otimes SO(6)\otimes \ SO(10)$& $1$\\
\textbf{}  & $b_2=  \{ \chi_{1256}\  \ y_{56}\  \ \omega_{12}\  \ \|\  \ \}$ & $I_{1256}\sigma_{56}$ &$\otimes \ E_8$&~\\
\cline{4-5}
\textbf{}  & $b_3=  \{   \ \|\   \tilde{\chi}_{3456}\  \  \ \tilde{\omega}_{3456}\ \}$ & $\bar{I}_{3456}$ &$SO(12) \otimes SO(16) \otimes E_8$ & 2\\
\textbf{}  & $b_4=  \{ \  \ \|\   \tilde{\chi}_{1256}\  \ \  \ \tilde{\omega}_{1256}\  \}$ & $\bar{I}_{1256}$ &~&~\\
\hline
\hline
\textbf{4}  & $b_1=  \{ \chi_{3456}\  \ y_{12}\  \ \omega_{123456}\  \ \|\   \tilde{y}_{123456}\  \ \tilde{\omega}_{123456}\ \}$ & $I_{3456}\sigma_{12}\bar{\sigma}_{123456}$ &$SO(2)^6\otimes SO(6)\otimes \ SO(10)$& $1$\\
\textbf{}  & $b_2=  \{ \chi_{1256}\  \ y_{56}\  \ \omega_{12}\  \ \|\  \ \}$ & $I_{1256}\sigma_{56}$ &$\otimes  \ E_8$&~\\
\cline{4-5}
\textbf{}  & $b_3=  \{ y_{123456}\  \ \omega_{123456}\  \ \|\   \tilde{\chi}_{3456}\  \ \tilde{y}_{12}\  \ \tilde{\omega}_{123456}\ \}$ & $\sigma_{123456}\bar{I}_{3456}\bar{\sigma}_{12}$ &$SO(4)^2\otimes SO(8)\otimes SO(12)$&2\\
\textbf{}  & $b_4=  \{\  \ \|\   \tilde{\chi}_{1256}\  \ \tilde{y}_{56}\  \ \tilde{\omega}_{12}\  \}$ & $\bar{I}_{1256}\bar{\sigma}_{56}$ &$\otimes \ E_8$&~\\
\hline
\hline
\textbf{5}  & $b_1=  \{ \chi_{3456}\  \ y_{12}\  \ \omega_{123456}\  \ \|\   \tilde{y}_{34}\  \ \tilde{\omega}_{34}\ \}$ & $I_{3456}\sigma_{12}\bar{\sigma}_{34}$ &$SO(2)^2\otimes SO(4)^2\otimes SO(6)$& $1$\\
\textbf{}  & $b_2=  \{ \chi_{1256}\  \ y_{56}\  \ \omega_{12}\  \ \|\  \ \}$ & $I_{1256}\sigma_{56}$ &$\otimes \ SO(10)\otimes  E_8$& ~\\
\cline{4-5}
\textbf{}  & $b_3=  \{ y_{34}\  \ \omega_{34}\  \ \|\   \tilde{\chi}_{3456}\  \ \tilde{y}_{12}\  \ \tilde{\omega}_{123456}\ \}$ & $\sigma_{34}\bar{I}_{3456}\bar{\sigma}_{12}$ &$SO(2)^2\otimes SO(10) \otimes SO(14)$& 2\\
\textbf{}  & $b_4=  \{   \ \|\   \tilde{\chi}_{1256}\  \ \tilde{y}_{56}\  \ \tilde{\omega}_{12}\   \}$ & $\bar{I}_{1256}\bar{\sigma}_{56}$ & $\otimes \ E_8$ &~\\
\hline
\hline
\textbf{6}  & $b_1=  \{ \chi_{3456}\ \ \omega_{3456}\  \ \|\  \tilde{y}_{1256}\  \ \tilde{\omega}_{1256}  \}$ & $I_{3456}\bar{\sigma}_{1256}$ &$SO(2)^4\otimes SO(4)^2\otimes \ SO(12)$& $1$\\
\textbf{}  & $b_2=  \{ \chi_{1256}  \ \omega_{1256}\  \ \|\  \ \}$ & $I_{1256}$ &$\otimes \ E_8$&~\\
\cline{4-5}
\textbf{}  & $b_3=  \{ y_{1256}\  \ \omega_{1256}\  \ \|\   \tilde{\chi}_{3456}\  \  \ \tilde{\omega}_{3456}\ \}$ & $\sigma_{1256}\bar{I}_{3456}$ &$SO(2) \otimes SO(6)^2 \otimes SO(14)$&2\\
\textbf{}  & $b_4=  \{ \  \ \|\   \tilde{\chi}_{1256}\  \   \ \tilde{\omega}_{1256}\ \}$ & $\bar{I}_{1256}$ &$\otimes \ E_8$&~\\
\hline
\hline
\textbf{7}  & $b_1=  \{ \chi_{3456}\ \ \omega_{3456}\  \ \|\  \tilde{y}_{1256}\  \ \tilde{\omega}_{1256}  \}$ & $I_{3456}\bar{\sigma}_{1256}$ &$SO(2)^4\otimes SO(4)^2\otimes \ SO(12)$& $1$\\
\textbf{}  & $b_2=  \{ \chi_{1256}  \ \omega_{1256}\  \ \|\  \ \}$ & $I_{1256}$ &$\otimes \ E_8$&~\\
\cline{4-5}
\textbf{}  & $b_3=  \{ y_{34}\  \ \omega_{34}\  \ \|\   \tilde{\chi}_{3456}\  \tilde y_{12} \ \tilde{\omega}_{123456}\ \}$ & $\sigma_{34}\bar{I}_{3456}$ &$SO(2) \otimes SO(6)^2 \otimes SO(14)$&2\\
\textbf{} & $b_4=  \{ \  \ \|\   \tilde{\chi}_{1256}\  \   \ \tilde y_{56} \ \tilde{\omega}_{12}\ \}$ & $\bar{I}_{1256}\bar\sigma_{56}$ &$\otimes \ E_8$&~\\
\hline
\hline
\textbf{8}  & $b_1=  \{ \chi_{3456}\  \ y_{12}\  \ \omega_{123456}\  \ \|\   \tilde{y}_{34}\  \ \tilde{\omega}_{34}\ \}$ & $I_{3456}\sigma_{12}\bar{\sigma}_{34}$ &$SO(2)^3\otimes SO(4)^2\otimes \ SO(14)$& $1$\\
\textbf{}  & $b_2=  \{ \chi_{1256}\  \ y_{56}\  \ \omega_{12}\  \ \|\   \ \}$ & $I_{1256}\sigma_{56}$ &$\otimes \ E_8$&~\\
\cline{4-5}
\textbf{}  & $b_3=  \{ y_{1256}\  \ \omega_{1256}\ \|\   \tilde{\chi}_{3456}\  \ \omega_{3456}\ \}$ & $\sigma_{1256}\bar{I}_{3456}$ &$SO(2)^2 \otimes SO(6) \otimes SO(18)$& 2\\
\textbf{}  & $b_4=  \{   \ \|\   \tilde{\chi}_{1256}\  \ \tilde{\omega}_{1256}\  \}$ & $\bar{I}_{1256}$ &$\otimes \ E_8$&~\\
\hline
\hline
\textbf{9}  & $b_1=  \{\chi_{3456}\  \ y_{12}\  \ \omega_{123456}\  \ \|\   \tilde{y}_{12}\  \ \tilde{\omega}_{12}\ \}$ & $I_{3456}\sigma_{12}\bar{\sigma}_{12}$ &$SO(2)^3\otimes SO(4)$& $1$\\
\textbf{}  & $b_2=  \{\chi_{1256}\  \ y_{56}\  \ \omega_{12}\  \ \|\  \ \}$ & $I_{1256}\sigma_{56}$ &$SO(6) \otimes  SO(12)$&~\\
\cline{5-5}
\textbf{}  & $b_3=  \{y_{123456}\  \ \omega_{123456}\  \ \|\   \tilde{\chi}_{3456}\  \ \tilde{y}_{12}\  \ \tilde{\omega}_{123456}\ \}$ & $\sigma_{123456}\bar{I}_{3456}\bar{\sigma}_{12}$ &$\otimes \ E_8$&2\\
\textbf{}  & $b_4=  \{y_{34}\  \ \omega_{34}\  \ \|\   \tilde{\chi}_{1256}\  \ \tilde{y}_{34}\  \ \tilde{\omega}_{123456}\ \}$ & $\sigma_{34}\bar{I}_{1256}\bar{\sigma}_{34}$ &~&~\\
\hline
\hline
\textbf{10}  & $b_1=  \{ \chi_{3456}\  \ \omega_{3456}\  \ \|\    \tilde{y}_{3456}\  \ \tilde{\omega}_{3456} \}$ & $I_{3456}\bar{\sigma}_{3456}$ &$SO(2)^3\otimes SO(4)^2$& $1$\\
\textbf{}  & $b_2=  \{ \chi_{1256}  \ \omega_{1256}\  \ \|\ \ \}$ & $I_{1256}$ &$\otimes \ SO(10)$&~\\
\cline{5-5}
\textbf{}  & $b_3=  \{ y_{123456}\  \ \omega_{123456}\  \ \|\   \tilde{\chi}_{3456}\  \ \tilde{y}_{12}\  \ \tilde{\omega}_{123456}\ \}$ & $\sigma_{123456}\bar{I}_{3456}\bar{\sigma}_{12}$ &$\otimes \ E_8$&2\\
\textbf{}  & $b_4=  \{ y_{34}\  \ \omega_{34}\  \ \|\   \tilde{\chi}_{1256}\  \ \tilde{y}_{34}\  \ \tilde{\omega}_{123456}\ \}$ & $\sigma_{34}\bar{I}_{1256}\bar{\sigma}_{34}$ &~&~\\
\hline
\hline
\textbf{11}  & $b_1=  \{ \chi_{3456}\  \ y_{12}\  \ \omega_{123456}\  \ \|\   \tilde{y}_{123456}\  \ \tilde{\omega}_{123456}\ \}$ & $I_{3456}\sigma_{123456}\bar{\sigma}_{12}$ &$SO(2)^7\otimes SO(4)\otimes \ SO(10)$& $1$\\
\textbf{}  & $b_2=  \{ \chi_{1256}\  \ y_{34}\  \ \omega_{123456}\  \ \|\   \tilde{y}_{34}\  \ \tilde{\omega}_{34}\ \}$ & $I_{12}\sigma_{34}\bar{\sigma}_{34}$ &$\otimes \ E_8$&~\\
\cline{4-4}
\textbf{}  & $b_3=  \{ y_{12}\  \ \omega_{12}\  \ \|\   \tilde{\chi}_{3456}\  \ \tilde{y}_{12}\  \ \tilde{\omega}_{123456}\ \}$ & $\sigma_{12}\bar{I}_{3456}\bar{\sigma}_{12}$ &$SO(2)^3\otimes SO(4) \otimes SO(6)$&~\\
\textbf{} & $b_4=  \{   \ \|\   \tilde{\chi}_{1256}\  \ \tilde{y}_{56}\  \ \tilde{\omega}_{12}\  \}$ & $\bar{I}_{1256}\bar{\sigma}_{56}$ &$ \otimes \ SO(12)\otimes  E_8$&~\\
\hline
\hline
\textbf{12}  & $b_1=  \{ \chi_{3456}\  \ y_{12}\  \ \omega_{123456}\  \ \|\   \tilde{y}_{123456}\  \ \tilde{\omega}_{123456}\ \}$ & $I_{3456}\sigma_{123456}\bar{\sigma}_{12}$ &$SO(2)^7\otimes SO(4)\otimes \ SO(10)$& $1$\\
\textbf{} & $b_2=  \{ \chi_{1256}\  \ y_{34}\  \ \omega_{123456}\  \ \|\   \tilde{y}_{34}\  \ \tilde{\omega}_{34}\ \}$ & $I_{12}\sigma_{34}\bar{\sigma}_{12}$ &$\otimes \ E_8$&~\\
\cline{4-4}
\textbf{}  & $b_3=  \{ y_{3456}\  \ \omega_{3456}\  \ \|\   \tilde{\chi}_{3456}\  \   \ \tilde{\omega}_{3456}\ \}$ & $\sigma_{3456}\bar{I}_{3456}$ &$SO(2)^3\otimes SO(4) \otimes SO(6)$&~\\
\textbf{} & $b_4=  \{   \ \|\   \tilde{\chi}_{1256}\  \   \ \tilde{\omega}_{1256}\  \}$ & $\bar{I}_{1256}$ &$\otimes \ SO(12) \otimes  E_8$ &~\\
\hline
\hline
\textbf{13}  & $b_1=  \{ \chi_{3456}\  \ y_{12}\  \ \omega_{123456}\  \ \|\   \tilde{y}_{123456}\  \ \tilde{\omega}_{123456}\ \}$ & $I_{3456}\sigma_{123456}\bar{\sigma}_{12}$ &$SO(2)^7\otimes SO(4)$& $1$\\
\textbf{} & $b_2=  \{ \chi_{1256}\  \ y_{34}\  \ \omega_{123456}\  \ \|\   \tilde{y}_{12}\  \ \tilde{\omega}_{12}\ \}$ & $I_{12}\sigma_{34}\bar{\sigma}_{12}$ &$\otimes \ SO(10)$&~\\
\textbf{}  & $b_3=  \{ y_{123456}\  \ \omega_{123456}\  \ \|\   \tilde{\chi}_{3456}\  \ \tilde{y}_{12}\  \ \tilde{\omega}_{123456}\ \}$ & $\sigma_{123456}\bar{I}_{3456}\bar{\sigma}_{12}$ &$\otimes \ E_8$&~\\
\textbf{} & $b_4=  \{ y_{12}\  \ \omega_{12}\  \ \|\   \tilde{\chi}_{1256}\  \ \tilde{y}_{34}\  \ \tilde{\omega}_{123456}\  \}$ & $\sigma_{12}\bar{I}_{1256}\bar{\sigma}_{34}$ &~&~\\
\hline
\hline
\textbf{14} & $b_1=  \{\chi_{3456}\ \ y_{12}\  \ \omega_{123456}\  \ \|\   \tilde{\tilde{y}}_{12}\  \ \tilde{\omega}_{12}\ \}$ & $I_{3456}\sigma_{12}\bar{\sigma}_{12}$ &$SO(2)^5\otimes SO(4)^2$& $1$\\
\textbf{}  & $b_2=  \{\chi_{1256}\  \ y_{34}\  \ \omega_{123456}\  \ \|\   \tilde{\tilde{y}}_{56}\  \ \tilde{\omega}_{56}\ \}$ & $I_{1256}\sigma_{34}\bar{\sigma}_{56}$ &$\otimes \ SO(10)$&~\\
\textbf{} & $b_3=  \{y_{12}\  \ \omega_{12}\  \ \| \   \tilde{\chi}_{3456}\  \ \tilde{y}_{12}\  \ \tilde{\omega}_{123456}\ \}$ & $ \sigma_{12} \bar{I}_{3456} \bar{\sigma}_{12}$ &$\otimes \ E_8$&~\\
\textbf{}  & $b_4=  \{y_{56}\  \ \omega_{56}\  \ \| \   \tilde{\chi}_{1256}\  \ \tilde{y}_{34}\  \ \tilde{\omega}_{123456}\ \}$ & $ \sigma_{56} \bar{I}_{1256} \bar{\sigma}_{34}$ &~&~\\
\hline
\hline
\textbf{15}  & $b_1=  \{\chi_{3456} \ y_{12} \ \omega_{123456} \ \| \ \tilde{y}_{12} \ \tilde{\omega}_{12} \}$ & $I_{3456}\sigma_{12}\bar{\sigma}_{12}$ & $SO(2)^6\otimes SO(4)$& $1$\\
\textbf{} & $b_2=  \{\chi_{1256} \ y_{34} \ \omega_{123456} \ \| \ \tilde{y}_{34} \ \tilde{\omega}_{34} \}$ & $I_{1256}\sigma_{34}\bar{\sigma}_{34}$ &$\otimes \ SO(12)$&~\\
\textbf{} & $b_3=  \{y_{12} \ \omega_{12} \ \|  \ \tilde{\chi}_{3456} \ \tilde{y}_{12}  \ \tilde{\omega}_{123456} \}$ & $ \sigma_{12} \bar{I}_{3456} \bar{\sigma}_{12}$ &$\otimes \ E_8$&~\\
\textbf{} & $b_4=  \{y_{34} \ \omega_{34} \ \| \ \tilde{\chi}_{1256} \ \tilde{y}_{34} \ \tilde{\omega}_{123456} \}$ & $\sigma_{34} \bar{I}_{1256} \bar{\sigma}_{34}$ &~&~\\
\hline
\hline
\textbf{16} & $b_1=  \{\chi_{3456} \ y_{12} \ \omega_{123456} \ \| \ \tilde{y}_{12} \ \tilde{\omega}_{12} \}$ & $I_{3456}\sigma_{12}\bar{\sigma}_{12}$ & $SO(2)^6\otimes SO(4)$& $1$\\
\textbf{}  & $b_2=  \{\chi_{1256} \ y_{34} \ \omega_{123456} \ \| \ \tilde{y}_{34} \ \tilde{\omega}_{34} \}$ & $I_{1256}\sigma_{34}\bar{\sigma}_{34}$ &$\otimes \ SO(12)$&~\\
\textbf{} & $b_3=  \{y_{3456} \ \omega_{3456} \ \|  \ \tilde{\chi}_{3456} \  \ \tilde{\omega}_{3456} \}$ & $ \sigma_{3456} \bar{I}_{3456} $ &$\otimes \ E_8$&~\\
\textbf{}  & $b_4=  \{y_{1256} \ \omega_{1256} \ \| \ \tilde{\chi}_{1256} \  \ \tilde{\omega}_{1256} \}$ & $\sigma_{1256} \bar{I}_{1256} $ &~&~\\
\hline
\hline
\textbf{17}  & $b_1=  \{ \chi_{3456}\  \omega_{3456}\  \ \|\   \tilde{y}_{3456}\  \ \tilde{\omega}_{3456}\   \}$ & $I_{3456}\bar{\sigma}_{3456}$ &$SO(2)^9\otimes SO(10)$& $1$\\
\textbf{}  & $b_2=  \{\chi_{1256}\  \omega_{1256}\  \ \|\   \tilde{y}_{1256}\  \ \tilde{\omega}_{1256}\   \}$ & $I_{1256}\bar{\sigma}_{1256}$ &$\otimes \ E_8$&~\\
\textbf{} & $b_3=  \{ y_{12}\  \ \omega_{12}\  \ \|\   \tilde{\chi}_{3456}\  \ \tilde{y}_{12}\  \ \tilde{\omega}_{123456}\ \}$ & $\sigma_{12}\bar{I}_{3456}\bar{\sigma}_{12}$ &~&~\\
\textbf{}  & $b_4=  \{ y_{34}\  \ \omega_{34}\  \ \|\   \tilde{\chi}_{1256}\  \ \tilde{y}_{34}\  \ \tilde{\omega}_{123456}\ \}$ & $\sigma_{34}\bar{I}_{1256}\bar{\sigma}_{34}$ &~&~\\
\hline
\hline
\textbf{18} & $b_1=  \{ \chi_{3456}\  \omega_{3456}\  \ \|\   \tilde{y}_{3456}\  \ \tilde{\omega}_{3456}\   \}$ & $I_{3456}\bar{\sigma}_{3456}$ &$SO(2)^9\otimes SO(10)$& $1$\\
\textbf{}  & $b_2=  \{\chi_{1256}\  \omega_{1256}\  \ \|\   \tilde{y}_{1256}\  \ \tilde{\omega}_{1256}\   \}$ & $I_{1256}\bar{\sigma}_{1256}$ &$\otimes \ E_8$&~\\
\textbf{} & $b_3=  \{ y_{3456}\  \ \omega_{3456}\  \ \|\   \tilde{\chi}_{3456}\  \ \  \ \tilde{\omega}_{3456}\ \}$ & $\sigma_{3456}\bar{I}_{3456}$ &~&~\\
\textbf{}  & $b_4=  \{ y_{1256}\  \ \omega_{1256}\  \ \|\   \tilde{\chi}_{1256}\  \ \  \ \tilde{\omega}_{1256}\ \}$ & $\sigma_{1256}\bar{I}_{1256}$ &~&~\\

\end{longtable}}



\subsection{Parallel with type II models}

In  \cite{TfoldsI} 18 independent Type II models were also found. However many of these involved Ising fermions, which we have excluded here. Only 5 of those models did not require Ising fermions, because their setup was more restrictive: $b_3$ and $b_4$ were the exact mirrors of  $b_1$ and $b_2$, respectively. In our case only the twists are mirrored between $b_1$ and $b_3$ and between $b_2$ and $b_4$, while the shifts are only constrained by modular invariance.  Moreover, in \cite{TfoldsI} zero shifts were not allowed.
Out of our 18 models there are 8 cases where $b_3$ is the mirror of $b_1$ and $b_4$ is the mirror of $b_2$, models 1, 4, 5, 6, 13, 14, 15 and 18. However, models 1, 4, 5 and 6 have no equivalent in \cite{TfoldsI} because they contain zero shifts. This leaves models 13, 14, 15 and 18. Referring to Table 1 in \cite{TfoldsI} and to the labeling in our Table \ref{tabmodels} the parallel between the type II models and the heterotic ones goes as follows.
The model which has effective Hodge numbers (9,9) in type II corresponds to our model 14,
the (15,3) model  corresponds to our model 15 and
the (3,15) model  corresponds to our model 18.
The (5,17) and (17,5) models both correspond to our model 13. Model (5,17) is linked to model 13 by $LR$ \footnote{See Appendix \ref{sets} for the defintion of these symmetries.}. Model (17,5) is related to 13 by $B_2B_4$, after the redefinitions $y_6\leftrightarrow \omega_6$ and $\bar y_6\leftrightarrow \bar \omega_6$\footnote{There is a typo in Table 1 in \cite{TfoldsI} :  in model (17,5) $b_3$ contains an extra $\bar\sigma_6$. }.

In the Type II case, even in non-geometric constructions, the surviving $N=2$ space-time supersymmetry allows to define ``effective'' Hodge numbers from the neutral massless spectrum. Indeed, by counting the number of massless vector and hyper-multiplets, one can set $h^{eff}_{11} = n_H -1$, $h^{eff}_{21} = n_V$ in the Type IIB case and $h^{eff}_{11} = n_V$, $h^{eff}_{21} = n_H - 1$ in the Type IIA case.

In the heterotic case, the surviving $N=1$ space-time supersymmetry allows for charged chiral multiplets and it seems harder to define effective topological invariants of the putative vacuum gauge bundle $\cE$, particularly because we start from the enhanced gauge symmetry point $SO(28)\times E_8$. For models with $SO(10)$ gauge group embedded in one $E_8$ factor, $\cE$ is a stable holomorphic vector bundle of rank $r=4$ with $C_1(\cE)=0$ and $C_2(\cE)=C_2(T)$. Massless ${\bf 16}$ are associated to $H^*(M,\cE)$, massless ${\bf 10}$ to $H^*(M,\wedge^2 \cE)$, adjoints ${\bf 45}$ to $H^*(M,{\cal O})$, while singlets are associated to $H^*(M,End\cE)$. Chiral asymmetry is given by $n_{\bf 16} - n_{{\bf 16}^*} = {1\over 2} C_3(\cE)$ \cite{Andersonetal}. In the non-geometric setting, one can count the number of massless ${\bf 10}$,  ${\bf 16}$,  ${\bf 16}^*$, ${\bf 45}$,  but it is far from obvious that one could define any meaningful effective topological invaraints
$C^{eff}_3(\cE)$,  $H_{eff}^*(M,\cE)$, $H_{eff}^*(M,\wedge^2 \cE)$ or $H_{eff}^*(M,{\cal O})$. Even neutral moduli are hard to associate to ``effective'' cohomology classes since  in addition to the would-be geometric ones, to be counted by $h^{eff}_{11}$ or  $h^{eff}_{21}$, one finds a plethora of charged singlets that should be associated to a putative $H_{eff}^*(M,End\cE)$.

Although for generic non-geometric heterotic models the situation seems hopeless, for models with gauge group $G_H \supset SO(10)\times E_8$ one can apply the ``inverse'' Gepner map  (\ref{invegepmap}) and get a Type II model with at least $N=1$ supersymmetry. As already mentioned, if the charged heterotic spectrum assembles into complete representations of different $E_6$'s in different sectors, the resulting Type II model enjoys $N=2$ supersymmetry and one can define ``effective'' Hodge numbers $h_{11}^{eff} = n_H - 1, h_{21}^{eff} = n_V$, for Type IIB. A necessary but not sufficient condition for this to happen is $n_{\bf 10} = n_{\bf 16} + n_{{\bf 16}^*}$, which characterizes ``self-dual'' models under spinor-vector duality \cite{spinvectdual}. Finally one can go back to the massless heterotic spectrum and set
\be
n_{\bf 1} = 1+ h_{11}^{eff} + h_{21}^{eff} + h_1(End \cE)   \rightarrow  h_1(End \cE)=
n_ {\bf 1} - 1- h_{11}^{eff} - h_{21}^{eff}  \ee
where $n_ {\bf 1}$ means the total number of $SO(10)$ singlets, including states
charged with respect to various $U(1)$'s or even non-abelian factors. Moreover 
\bea && n_{\bf 16} = \chi^{eff}(\cE) \qquad
n_{\bf 10} = \chi^{eff}(\wedge^2 \cE) \qquad
n_{\bf 45} = \chi^{eff}({\cal O}) \nn \\
&& n_{{\bf 16}^*} = n_{\bf 16}  - {1\over 2} C_3^{eff}(\cE) = \chi^{eff}(\cE) - {1\over 2}
C_3^{eff}(\cE) \eea
where $\chi^{eff} (\cE) = \sum_k (-)^k h^{k,0}(\cE)$ denotes the generalized arithmetic genus. When the heterotic spectrum cannot be assigned to complete $E_6$ representations, one can still try and define ``effective'' Hodge numbers, in particular  $h_1(End\cE) = n^{Het}_ {\bf 1} - n^{II}_ {\bf 1}$,  but the geometric meaning is not at all clear. Among our models with an $SO(10)$ gauge group, including their discrete torsion variants, there are no self-dual examples.

Finally, it would be interesting on the one hand to construct Type II analogues of some of the L-R asymmetric heterotic string models with non mirror basis sets, and on the other hand to extend our present analysis allowing for Ising fermions.  The heterotic string counterpart of models like the ``minimal'' Type II with  $h^{eff}_{11} =h^{eff}_{21} =1$, should have the same neutral ``geometric'' moduli, but also many (possibly charged) singlets, and {\it a priori} should be non-chiral. Aside from the last draw-back, they would probably share some similarities
with the heterotic models recently constructed in \cite{Andersonetal}, at least for the low-energy effective supergravity properties.

 In some cases, \ie when the Type II modular invariant partition function consists of several disjoint orbits of the modular group, one can apply different Gepner maps in different sectors and get  chiral heterotic models even if  $h_{11}^{eff}=h_{21}^{eff}$ to start with. This relies on the ambiguity of mapping the characters $-S_2$ and $-C_2$ into $S_{10}$ and $C_{10}$ or {\it vice versa}. Usually Gepner map associates heterotic K\"ahler deformations plus chiral multiplets in the ${\bf 27}$ to Type II K\"ahler deformations  and heterotic complex deformations plus chiral multiplets in the ${\bf 27}^*$ to Type II complex deformations, so that $N_{\bf 27} - N_{{\bf 27}^*} = h_{11}^{eff}-h_{21}^{eff}$.  Yet, when disjoint modular orbits are present, which allow for the introduction of discrete torsion, one can associate heterotic K\"ahler deformations plus chiral multiplets in the ${\bf 27}^*$  to some Type II K\"ahler deformations  and heterotic complex  deformations plus chiral multiplets in the ${\bf 27}$ to some other Type II complex deformations. As a result  $N_{\bf 27} - N_{{\bf 27}^*} =( h_{11}^{eff} + n)- (h_{21}^{eff} -n) \neq h_{11}^{eff}-h_{21}^{eff}$. We will see this mechanism at work in some explicit example in Appendix C, where the role of discrete torsion is analyzed in some detail. 

\section{Moduli in free fermionic models}

 Moduli in the free fermionic models have been discussed extensively in \cite{Lopez:1994ej}. For completeness we include here a review of some relevant aspects and later on discuss exactly marginal charged deformations that ``higgs'' the non-abelian gauge group.

 \subsection{\{F, S\} model}

 Consider the model (\ref{FS}) generated by the set $\{F, S\}$. The gauge bosons of $SO(44)$, arising from the untwisted Neveu-Schwarz sector, can be expressed in terms of 22 right-moving complex fermions $\tilde\Psi^{+M}$ and their complex conjugates $\tilde\Psi^{-M}$
  \bea &&\left|\psi^\mu\right> \otimes \left |\tilde\Psi^{+M}\tilde\Psi^{-M} \right>\ , \quad M,N=1,...,22 \ ; \label{cartan}\\
&&\left|\psi^\mu\right> \otimes \left|\tilde\Psi^{\pm M} \tilde\Psi^{\pm N}\right>\ , \quad M > N\ .\eea
The first line describes the Cartan subalgebra, while the second corresponds to the non-zero roots of
$SO(44)$. The massless scalar states are also part of the NS sector and transform in the adjoint representation
\bea
&&\left|\chi^I\right>  \otimes\left |\tilde\Psi^{+M} \tilde\Psi^{-M} \right> \ ,\quad   M=1...22\ , \  I=1...6 \ ;\label{moduli}\\
&&\left|\chi^I\right>\otimes  \left|\tilde\Psi^{\pm M}\tilde\Psi^{\pm N}\right>\ ,\quad  M,N=1...22\ , \  I=1...6 \ .\label{matter}\eea
Clearly, the 6$\times$22 states in the first line are also in the Cartan sub-algebra of $SO(44)$. They correspond to the geometrical moduli of this model: the background metric $G_{ij}$, the antisymmetric tensor $B_{ij}$ and Wilson lines $A_{ia}$ ($i,j=1,...,6,\ a=1,...,16$). The states in (\ref{matter}) are the matter fields. In the eventuality of a rank reduction, some of them might become uncharged as well, if the corresponding Cartan generators in (\ref{cartan}) are projected out.

  The moduli (\ref{moduli}) can be rewritten as
  \bea G_{ij},\ B_{ij}: &\left|\chi^i\right>  \otimes \left |\tilde y^{j} \tilde\omega^{j} \right> \ ,   \  &i,j=1,...,6 \ ;\label{gb}\\
A_{ia}: &\left|\chi^i\right> \otimes \left|\tilde\Psi^{+ a}\tilde\Psi^{- a}\right>\ , \ &a=1,...,16\ .\label{wl} \eea
Because the boundary conditions of the worldsheet fermions $\chi^i$ are related by (\ref{susy}) to the ones of $y^i,\, \omega^i$, left-right asymmetric reflections of $\{y^i,\, \omega^i|\, \tilde y^i,\, \tilde\omega^i\}$ project out all the moduli in (\ref{gb}).

\subsection{NAHE model}
The massless non-chiral spectrum in the untwisted visible sector of the NAHE model consists of chiral multiplets in $({\bf 10}, {\bf 6}_I)$  and $({\bf 6}_J, {\bf 6}_K)$ with $J\neq K$ and  $I,J,K=1,2,3$. Denoting the former by $A^I_{a,i_I}$, with $a=1,...,10$ and $i_I = 1,...,6$, and the latter by $B^I_{i_J,i_K}$, the tri-linear (tree-level exact) superpotential\footnote{In toroidal orbifolds, the untwisted super-potential is always a truncation of the parent trilinear superpotential $W_{N=4} = \epsilon_{IJK} Tr(\phi^I [\phi^J,\phi^K])$.}
 inherited from $N=4$ reads
\be
W_{u-u-u} = |\varepsilon_{IJK}| \left(A^I_{a,i_I} B^{J,i_I}{}_{j_K}A^{K,j_K,a} +
B^{I,i_J}{}_{j_K} B^{J,j_K}{}_{k_I} B^{K,k_I}{}_{i_J}\right) \ .
\ee
In each of the three twisted sectors one gets a massless chiral spectrum with $S^{I,\Lambda}_{f_I,A_I}$ in the $({\bf 2}_{L,I}|{\bf 16}, {\bf 6}_I)$ and
$C^{I,\Lambda}_{\dot{f}_I,\bar{A}_I}$ in the $({\bf 2}_{R,I}|{\bf 16}^*, {\bf 6}_I)$.  The tri-linear u-t-t superpotential is of the form
\be
W_{u-t-t} = A^I_{a,i_I} \Gamma^a_{\Lambda\Sigma} \left[
 \Gamma^{i_I}_{A_IB_I} \varepsilon^{f_Ih_I} S^{I,\Lambda}_{f_I,A_I}S^{I,\Sigma}_{h_I,B_I} + \Gamma^{i_I}_{\bar{A}_I\bar{B}_I} \varepsilon^{\dot{f}_I\dot{h}_I} C^{I,\Lambda}_{\dot{f}_I,\bar{A}_I}C^{I,\Sigma}_{\dot{h}_I,\bar{B}_I}\right] \ .
 \nonumber
 \ee
Although, at first look, the model seems to evade the problem between moduli stabilization and chirality, since, except for the axion-dilaton, all massless fields are charged, at a closer inspection one can identify exact flat directions.
Setting for simplicity the twisted fields $S$ and $C$ to zero, the left-over superpotential is a truncation of the one for $N=4$ SYM. For this reason we do not expect higher order terms at tree level or perturbatively.
In turn the $N=4$ SYM admits flat directions for fields along the Cartan of the parent $SO(28)$. The fields of this kind surviving the $\Z_2\times \Z_2$ projection are exact flat directions. In particular one can break any pair of $SO(6)$ factors to the diagonal or $SO(6)\times SO(10)$ to $SO(6)_{diag} \times SO(4)$.
Further evidence for the existence of exactly marginal deformations is the existence of models with different fermionic sets and smaller symmetry.
The $\Z_2 \times \Z_2$ projections reduce the content of (\ref{gb}) and (\ref{wl}) to 12 moduli
\bea \left|\chi^I\right>  \otimes \left |\tilde y^{J} \tilde\omega^{J} \right>\ ,\quad I,J =\{1,2\},\{3,4\},\{5,6\}\ . \eea
They correspond to those states in the $(6_I, 6_J)$ representations that are part of the Cartan subalgebra.

\subsection{Heterotic counterpart of brane recombination}

The effect of turning on VEV's for scalar fields in the bi-fundamental representation is the heterotic counterpart of brane recombination \cite{ACDP, brarec2} in theories with open strings. From the worldsheet viewpoint two isomorphic current algebras at level one combine to give a single current algebra at level two. From the effective field theory viewpoint the product gauge group gets broken to the diagonal. In the NAHE model, for instance, one can break any product $SO(6)\times SO(6)$ to $SO(6)_{diag}$. One can also break $SO(10)\times SO(6)$ to $SO(4)\times SO(6)_{diag}$ where the first factor arises from the decomposition $SO(10)\rightarrow SO(4)\times SO(6)$.

More explicitly, one has to decompose the spectrum in representations of the current algebra at level two\footnote{See, for example, \cite{Ginsparg}.}. The ``coset'' CFT $SO(2n)_1 \times SO(2n)_1 / SO(2n)_2$ is the $\Z_2$-orbifold theory at $R^2 = 2n$, in units where $\alpha^\prime = 2$. Indeed, the central charge of $SO(2n)_{k}$ is $c^{(2n)}_{k} = {k} n(2n-1)/( {k} + 2n -2)$, so that $c^{(2n)}_2 = 2 c^{(2n)}_1 - 1$. The deficit is exactly compensated by a $c=1$ rational orbifold CFT. The spectrum of the latter consists of primaries of dimension $h=p^2/(4n)$ with $p=0,1,... n$, two spin fields of dimension $h=1/16$ and $h=9/16$ and a chiral ``current'' of dimension $h=1$ broken by the boundary conditions.

While at level ${k} =1$ the only ``integrable'' representations are the singlet $O_{2n}$, the vector $V_{2n}$ and the two spinors $S_{2n}$ and $C_{2n}$, at level ${k}=2$ new representations become integrable in that they satisfy the constraint $\rho\cdot w_R \le {k}$ where $\rho$ is the maximal root and $w_R$ is the highest weight of the representation $R$.

The scaling dimension of the primary in the representation $R$ is given by
$ h_{R,{k}}^{(2n)} = C_2(R) /( {k} + 2n - 2)$. In particular, at ${k}=2$ for the vector one finds
$ h_{V,2}^{(2n)} = (2n-1) / (4n) $, while for the adjoint $ h_{A,2}^{(2n)} = (n-1) / n$. It is amusing that for the symmetric traceless tensor one has $ h_{T,2}^{(2n)} = 1$, independently of $n$.

By current algebra analysis, one can derive the following decompositions
\bea
&&O_{2n} O_{2n} = \hat{O}_{2n} \xi_0 + \hat{A}_{2n} \xi_{1/n} + \hat{T}_{2n} \xi_{1} + ...  \ ;\nn \\
&&V_{2n} V_{2n} = \hat{O}_{2n} \xi_1 + \hat{A}_{2n} \xi_{1/n} + \hat{T}_{2n} \xi_{0} + ...  \ ;\nn \\
&&O_{2n} V_{2n} = \hat{V}_{2n} \xi_{1/4n} + ...  \ ;\nn \\
&&O_{2n} S_{2n} = \hat{S}_{2n} \xi_{1/16} + \hat{S}^\prime_{2n} \xi_{9/16} \ ; \nn \\
&&V_{2n} S_{2n} = \hat{C}_{2n} \xi_{9/16} + \hat{C}^\prime_{2n} \xi_{1/16} \ ,
\eea
where $\hat{L}_{2n}$ denote characters of $SO(2n)$ at ${k} =2$ and $\xi_{h}$ denote characters of the $c=1$ coset CFT.
Tensor products of spinorial representations depend on the parity of $n$. For $n$ even
one gets \bea
&&S_{2n} S_{2n} = \hat{O}_{2n} \xi_{n/4} + \hat{A}_{2n} \xi_{(n-2)^2/4n} + ...  \ ;\nn \\
&&S_{2n} C_{2n} = \hat{V}_{2n} \xi_{(n-1)^2/4n} + ...\ .
\eea
For $n$ odd one gets instead
\bea
&&S_{2n} S_{2n} = \hat{V}_{2n} \xi_{(n-1)^2/4n} + ... \ ,\nn \\
&&S_{2n} C_{2n} = \hat{O}_{2n} \xi_{n/4} + \hat{A}_{2n} \xi_{(n-2)^2/4n} + ...\ .
\eea
For our considerations it is crucial that $V_{2n} V_{2n}$ contains $\hat{O}_{2n} \xi_1$, \ie the combination of the singlet of $SO(2n)$ at ${k} =2$ and the ``current'' $\xi_1$. The latter corresponds to the (chiral) deformation of the radius under which all the primaries, except for the twist fields, change conformal dimension and $\xi_{h}$ decomposes into an infinite number of primaries of the resulting irrational CFT.
In order to get a modular invariant partition function one has to combine it with an anti-chiral deformation of the ``radius''. This is precisely what the anti-chiral part of the vertex operator for the bi-fundamental moduli fields does. Indeed in the non-canonical $q=0$ picture at zero momentum
\be
V = \partial X \bar\partial X^\prime \ ,
\ee
where $\bar\partial X^\prime$ represents the ``current'' in the orbifold CFT.
It is straightforward, though rather tedious, to decompose the entire partition function in terms of ``orbifold'' characters. In so far as the massless spectrum is concerned, in addition to rank reduction, a number of massless states get masses. The only states that remain massless are the ones that involve the 'coset' primary with fixed dimension, \ie the identity, the current and the two twist fields. A detailed analysis is beyond the scope of the present investigation and will be presented elsewhere \cite{hetbranerecomb}.

\section{Conclusions and outlook}

Let us summarize our results and draw lines for further investigation.
In the framework of the free fermionic construction,  we have studied non-geometric ``compactifications'' of the heterotic superstring in D=4, with a small number of neutral moduli. We performed a systematic scan for models with four {\it a priori} left-right asymmetric $\Z_2$ projections that eliminate neutral moduli and we identified 18 classes,  together with variants resulting from discrete torsions. For simplicity, we have focused on models with world-sheet fermions twisted at least in pairs in each sector, excluding thus Ising fermions. We were able to compare five of our classes to Type II models obtained in a previous study \cite{TfoldsI}.

Contrary to Type II models, where $N=2$ supersymmetry allows to define ``effective'' Hodge numbers even in such a non-geometric setting, we have not found in general a simple
interpretation of the massless spectrum in terms of ``effective'' topological invariants. Our construction corresponds to NAHE-like models with non-standard embedding of the spin connection in the (enhanced) gauge group. We cannot exclude that at least the chiral spectrum be associated to the topology of (non-abelian) gauge bundles existing only for special choice of the moduli, that are consequently frozen.  When  the charged spectrum assembles into complete representations of different  $E_6$'s in different sectors, an ``inverse'' Gepner map could allow to associate an $N=2$ Type II model to the heterotic model and to define reliable ``effective'' topological invariants.

Finally, we discussed exactly marginal deformations along charged directions and identified
patterns of symmetry breaking where product gauge groups,
realized at level one, are broken to their diagonal at higher level.
We should thus conclude that the exclusion of truly neutral moduli does not prevent the presence of exactly marginal deformations along charged directions. Eventually, adjoint scalars may appear that could break the gauge group to abelian factors. If and how this could be get rid of by FI terms or otherwise remains to be seen.
Another promising direction of investigation would be to include Ising fermions and some complex twists and shifts. The road that leads to phenomenologically viable chiral models with few moduli is still long and it may require to include fluxes and non-perturbative (NS5-brane instanton) effects. The advantage of the heterotic string over all other descriptions is that world sheet instantons are automatically incorporated whenever an exact CFT description is available. The only fluxes one can turn on are in  the NS sector, and so are amenable to a world sheet description and, finally, even NS5-brane instantons admit a world sheet description, at least in some limits.

\section*{Acknowledgments}
It is a pleasure to thank P.~Anastasopoulos, C.~Angelantonj, C.~Bachas, A.~Faraggi, C.~Kounnas, J.~F.~Morales, R.~Richter, A.~N.~Schellekens and Y.~S.~Stanev  for interesting discussions. G.P. would like to thank the Theory Group of the PH Division of CERN for the kind hospitality while this work was being completed. This work was partially supported by the ERC Advanced Grant n.226455 ``Superfields'', by the Italian MIUR-PRIN contract 2009KHZKRX-007.
The work of C. T. is supported by the Marie-Heim V\"{o}gtlin program of the Swiss National Science Foundation and the University of Bern. The work of L.T. is supported by German Science Foundation (DFG) within the
Collaborative Research Center 676 "Particles, Strings and the early
universe" and by the Research Training Group RTG 1670
(Graduiertenkolleg 1670) at Hamburg University.

\appendix

\section{Appendix: The most general choice of sets}
\label{sets}

 In this section we derive all the inequivalent models that can be obtained from our basis vectors (\ref{ansatz1}).  For this purpose, it is useful to write them in the slightly different form
\bea
 &&b_1 =(b_{1L},b_{1R})= \{\, I_{3456}\, S_1\, |\,\tilde S_1\, \} = \{ (\chi \, \omega)^{3456}  \, (y\, \omega)^{i_1 i_2 \ldots }  | (\tilde y\, \tilde \omega)^{k_1 k_2 \ldots }  \} \ , \nonumber\\
 &&b_2 =  (b_{2L},b_{2R})=\{\, I_{1256}\, S_2 \, | \,\tilde S_2\, \} = \{(\chi\,  \omega)^{1256}  \, (y\, \omega)^{j_1 j_2 \ldots }  | (\tilde y\, \tilde \omega)^{ l_1 l_2 \ldots }   \} \ , \nonumber\\
 &&b_3 =(b_{3L},b_{3R})= \{\, S_3 \, |\,\bar I_{3456}\, \tilde S_3\, \}  = \{ ( y\,  \omega)^{k'_1 k'_2 \ldots }  | (\tilde\chi\, \tilde  \omega)^{3456}
  (\tilde y\, \tilde \omega)^{i'_1 i'_2 \ldots }     \} \ , \nonumber\\
 &&b_4  =(b_{4L},b_{4R})=\{\, S_4 \, |\, \bar I_{1256}\, \tilde S_4\, \} = \{ ( y\,  \omega)^{ l'_1 l'_2 \ldots } | (\tilde\chi\, \tilde  \omega)^{1256}
(\tilde y\, \tilde \omega)^{j'_1 j'_2 \ldots }  \} \  , \label{set}\eea
where $S_{1,2}, \, \tilde S_{1,2},\, S_{3,4}$ and $\tilde S_{3,4}$ are groups of shifts. Since we exclude Ising fermions from our models, the groups of fermion labels $\{12\}$, $\{34\}$, $\{56\}$ are unsplit in the following.

In oder to determine the inequivalent models we need to derive the inequivalent shifts, since the twists are already fixed. Let's first discuss the shifts that accompany the twists, i.e. $S_{1,2}$ and $\tilde S_{3,4}$.  Certain shifts are redundant when they follow twists. For instance, a shift $S_1$ in the $34$ directions in $b_1$ can be reabsorbed by redefining $y_{34}\leftrightarrow \omega_{34}$, without affecting $b_2$ (or $b_{3,4}$). The same holds for a shift $S_2=y_{12}\omega_{12}$ in $b_2$. A shift in the 56 directions can be absorbed if present in both $S_1$ and $S_2$. If a shift in the 56 directions  is present only in $S_2$, then performing
$y_{56}\leftrightarrow \omega_{56}$ will remove the shift from $S_2$ and will make it appear in $S_1$. We can pick the convention under which the shift in the 56-directions is always in $S_2$, if present at all. As such, the left moving part of $b_1$ and $b_2$ can take on the following values
\bea &&b_{1L} \in \left\{~ b_{1L}^{(1)}=\{\chi^{3456}~  \omega_{3456}\},~b_{1L}^{(2)}=\{\chi^{3456}~ y_{12}~\omega_{1...6}\}~\right\},\nonumber\\
&&b_{2L} \in \left\{~b_{2L}^{(1)}= \{\chi^{1256}~ \omega_{1256}\},~b_{2L}^{(2)}=\{\chi^{1256}~ y_{56}~ \omega_{12}\},~b_{2L}^{(3)}=\{\chi^{1256}~ y_{34}~\omega_{1...6}\}\right.,\nonumber\\
&&~~~~~~~~~~\left. b_{2L}^{(4)}=\{\chi^{1256}~  y_{3456}\omega_{1234}\}\right\}.~\label{choice}
\eea
Modular invariance of $b_1$ implies that $b_{1L}^{(1)}$ can only be combined in the right moving part with a 4-shift or no shift, $\tilde{S}_1\in \left\{ \emptyset ,\{\tilde y_{ijkl}~\tilde\omega_{ijkl}\}\right\}$, while $b_{1L}^{(2)}$ requires a 2-shift or a 6-shift, $\tilde{S}_1\in \left\{ \{\tilde y_{ij}~\tilde\omega_{ij}\},\{\tilde y_{1...6}~\tilde \omega_{1...6}\}\right\}$, with $ ij,kl\in\{\{12\},\{34\},\{56\}\}$. Hence there are two choices for the left part of $b_1$ and eight possible choices for the shifts in the right part, but the first modular invariance condition reduces by half the viable possibilities.

Similar arguments apply to the four choices of $b_{2L}$. At this point there are 8 choices for $b_1$ and 16 choices for $b_2$. To implement the next modular invariance constraint, $b_1\cdot b_2=0~ \rm{mod} ~4$, we need to look at the left part of $b_1$ and $b_2$, because the right part always contributes a multiple of 4. Then $b_{1L}^{(1)}$ is compatible with $b_{2L}^{(1,4)}$ and $b_{1L}^{(2)}$ with $b_{2L}^{(2,3)}$. Hence this modular invariance condition also cuts by  half the number of possibilities.

The analysis can be repeated for $b_3$ and $b_4$, yielding
 \bea &&b_{3}^{(1)} = \left\{~ \emptyset~ {\rm or}~ \{ y_{ijkl}~\omega_{ijkl}\} ~|~\tilde \chi^{3456}~  \tilde \omega_{3456}\right\} ~~\rm{with}  \nonumber\\
 && b_4\in \left\{  ~ b_4^{(1)}= \left\{    ~  \emptyset ~ {\rm or}~ \{ y_{ijkl}~\omega_{ijkl}\} ~ ~|~\tilde\chi^{1256}~\tilde\omega_{1256}\right\} \right.,  \nonumber\\
  && ~~~~~~~~~\left. b_4^{(2)}= \left \{    ~   \{y_{ij}\omega_{ij}\}~ {\rm or}~ \{ y_{1..6}~\omega_{1...6}\} ~ ~|~\tilde\chi^{1256}~\tilde y_{3456}~\tilde\omega_{1234}\right\}        ~ \right\}\nonumber\\
  && ~~~~~~  {\rm  or} \nonumber\\
  &&b_{3}^{(2)} = \left\{~   \{y_{ij}\omega_{ij}\}~ {\rm or}~ \{ y_{1..6}~\omega_{1...6}\} ~|~\tilde \chi^{3456}~  \tilde y_{12}~ \tilde \omega_{1...6}\right\} ~~\rm{with} \nonumber\\
   &&  b_4\in \left\{ ~ b_4^{(3)}= \left\{    ~   \emptyset~ {\rm or}~ \{ y_{ijkl}~\omega_{ijkl}\} ~ ~|~\tilde\chi^{1256}~\tilde y_{56}~\tilde\omega_{12}\right\} \right., \nonumber\\
 && ~~~~~~~~~ \left. b_4^{(4)}=  \left\{    ~   \{y_{ij}\omega_{ij}\}~ {\rm or}~ \{ y_{1..6}~\omega_{1...6}\} ~ ~|~\tilde\chi^{1256}~\tilde y_{34}~\tilde\omega_{1...6}\right\}        ~ \right\}.\label{choice2}
   \eea
 Thus there are $2^{12}$ different modular invariant models, many of which are still equivalent. The other modular invariant constraints should reduce this number to the $2^8=256$ different models found by the algorithm in appendix E, hence a reduction by a factor of  $2^4$, corresponding to the four modular invariance conditions related to the products $b_{1,2}\cdot b_{3,4}$.

  Next we search the inequivalent values for the shifts $\tilde S_{1,2}$ and $S_{3,4}$. For this we remark that a given set $\{b_1, b_2, b_3,b_4\}$  carries certain symmetries that transform it into an equivalent set of the form (\ref{set}). These symmetries can be expressed as
 \bea &&B_1 : b_1\rightarrow b_1+b_2 ~, ~ \{\chi_{12}~ y_{12} ~\omega_{12}\}\leftrightarrow \{\chi_{56} ~y_{56} ~\omega_{56}\},\nonumber\\
  &&B_2 : b_2\rightarrow b_1+b_2 ~, ~ \{\chi_{34}~ y_{34} ~\omega_{34}\}\leftrightarrow \{\chi_{56} ~y_{56} ~\omega_{56}\},\nonumber\\
 &&B_3 : b_3\rightarrow b_3+b_4 ~, ~ \{\tilde\chi_{12}~ \tilde y_{12} ~\tilde\omega_{12}\}\leftrightarrow \{\tilde\chi_{56} ~\tilde y_{56} ~\tilde\omega_{56}\},\nonumber\\
 &&B_4 : b_4\rightarrow b_3+b_4 ~, ~ \{\tilde\chi_{34}~\tilde y_{34} ~\tilde\omega_{34}\}\leftrightarrow \{\tilde\chi_{56} ~\tilde y_{56} ~\tilde\omega_{56}\},\nonumber\\
 &&L :  \{\chi_{12}~ y_{12} ~\omega_{12}\}\leftrightarrow \{\chi_{34} ~y_{34} ~\omega_{34}\},\nonumber\\
  &&R :  \{\tilde\chi_{12}~\tilde y_{12} ~\tilde\omega_{12}\}\leftrightarrow \{\tilde\chi_{34} ~\tilde y_{34} ~\tilde\omega_{34}\}.\label{sym}
 \eea
The transformation $B_1$ adds the shifts  $\tilde S_2$ from the basis set $b_{2}$ to the shifts  $\tilde S_1$ in $b_{1}$, while $B_2$ adds the shifts from $b_{1}$ to the ones in $b_{2}$ (and similar for $B_{3,4}$). The transformation $L=B_1B_2B_1=B_2B_1B_2$ is basically switching the shifts $\tilde{S_1}$ and $\tilde{S_2}$ between $b_1$ and $b_2$ and, likewise, $R=B_3B_4B_3=B_4B_3B_4$ is switching $S_3$ and $S_4$. In order to always come back to  our conventional choice in (\ref{choice}) and (\ref{choice2}), one also has to reabsorb some shifts in $S_{1,2}$ and $\tilde S_{3,4}$ as explained in the previous appendix. This implies that sometimes the transformations in (\ref{sym}) are accompanied by redefinitions of the form $y_{ij}\leftrightarrow \omega_{ij}$ or $\tilde y_{ij}\leftrightarrow \tilde \omega_{ij}$, with $ij\in \{ \{12\}, \{34\}, \{56\}\}$.


The transformations in (\ref{sym}) form a group of 36 elements with identity :
$$\begin{array}{ccccccccc}
 I & B_1  & B_2  & B_3 &B_4 &B_1B_3 &B_1B_4 &B_2B_3 &B_2B_4\\
 L &B_1L &B_2L &B_3L &B_4L &B_1B_3L &B_1B_4L &B_2B_3L &B_2B_4L\\
 R &B_1R &B_2R &B_3R &B_4R &B_1B_3R &B_1B_4R &B_2B_3R &B_2B_4R\\
 LR &B_1LR &B_2LR &B_3LR &B_4LR &B_1B_3LR &B_1B_4LR &B_2B_3LR &B_2B_4LR.
\end{array}$$
Hence, each class of equivalent models should contain a priori 36 elements. However, in most cases, some of the previous symmetries are trivial due to the particular form of the basis vectors(for instance lack of shifts), so some classes will contain fewer elements.

For the shifts $\tilde S_{1,2}$ the relevant symmetries are $B_1, \, B_2$ and $L$. The possible values of $\left(\begin{array}{c} \tilde S_1 \\ \tilde S_2 \end{array}\right)$ split into the following equivalence classes according to these three symmetries
 \begin{enumerate}
 \item[$\bullet$] Both zero,  $\left(\begin{array}{c} \tilde S_1 \\ \tilde S_2 \end{array}\right)=\left(\begin{array}{c} 0 \\ 0 \end{array}\right)$. The symmetries $B_1, \, B_2, \, L$ are trivial in this case.

 \item[$\bullet$] One zero,  $\left(\begin{array}{c} \tilde S_1 \\ \tilde S_2 \end{array}\right)=\left(\begin{array}{c} x \\ 0 \end{array}\right)$,  with $x$ being a 2-shift ($y_{ij}\omega_{ij}$), a 4-shift ($y_{ijkl}\omega_{ijkl}$) or 6-shift ($y_{1...6}\omega_{1...6}$). The case $\left(\begin{array}{c} x \\ 0 \end{array}\right)$ is equivalent through $L$ to $\left(\begin{array}{c} 0 \\ x \end{array}\right)$ and by $B_2$ to $\left(\begin{array}{c} x \\ x \end{array}\right)$. $B_1$ is trivial in this context.

\item[$\bullet$] Two different 2-shifts\footnote{The case of two equal 2-shifts is contained in the previous class.}, $\left(\begin{array}{c} 2_x \\ 2_y \end{array}\right)$, $x\neq y$. By $L$ this is equivalent to $\left(\begin{array}{c} 2_y \\ 2_x \end{array}\right)$, while $B_1$ and $B_2$, alone or combined with $L$, yield $\left(\begin{array}{c} 2_{x ~or~ y} \\ 4_{xy} \end{array}\right)$ and $\left(\begin{array}{c} 4_{xy} \\ 2_{x ~or ~y} \end{array}\right)$.

\item[$\bullet$] $\left(\begin{array}{c} 4_{xy} \\ 4_{xz} \end{array}\right)$, $y \neq z$, together with $\left(\begin{array}{c} 4_{xz} \\ 4_{xy} \end{array}\right)$, $\left(\begin{array}{c} 4_{xy} \\ 4_{yz} \end{array}\right)$, $\left(\begin{array}{c} 4_{yz} \\ 4_{xz} \end{array}\right)$, $\left(\begin{array}{c} 4_{xz} \\ 4_{yz} \end{array}\right)$, $\left(\begin{array}{c} 4_{yz} \\ 4_{xy} \end{array}\right)$.

 \item[$\bullet$] $\left(\begin{array}{c} 6 \\ 2_x \end{array}\right)$, together with  $\left(\begin{array}{c} 2_x \\ 6 \end{array}\right)$,  $\left(\begin{array}{c} 6 \\ 4_{yz} \end{array}\right)$, $\left(\begin{array}{c} 4_{yz} \\ 6 \end{array}\right)$, $\left(\begin{array}{c} 4_{yz} \\ 2_x \end{array}\right)$ and $\left(\begin{array}{c} 2_x \\ 4_{yz} \end{array}\right)$.

  \end{enumerate}

  Due to the symmetry of our setup the values of $ S_{3,4}$ fall into the same equivalence classes as $\tilde S_{1,2}$.  We can think of the values listed above as building blocks. The last step we need to take is to determine how the values of $\tilde S_{1,2}$ combine with those of $S_{3,4}$ in order to respect modular invariance.  We report in the appendix B the resulting models.

  \section{Appendix: Inequivalent models}
  \label{ineq}

  In this section we combine the various possible values of $\tilde S_{1,2} $ with those of $S_{3,4}$, keeping in mind the restrictions and conventions in (\ref{choice}), (\ref{choice2}), in order to write all possible inequivalent models,  gathered in Table  \ref{tabmodels}\footnote{The order of the models in the table is slightly altered in respect to this section, because it takes into account repetitions of the gauge group.}.

  The first option is to put to zero all four shifts $ \tilde S_{1,2},\,  S_{3,4}$ .  This leads to the first model in Table \ref{tabmodels}. All symmetries (\ref{sym}) are trivial in this case.

   Next we combine the building block $\left(\begin{array}{c}  S_3 \\  S_4 \end{array}\right)=\left(\begin{array}{c}  0 \\  0 \end{array}\right)$ with one of the others.
   The form of the  basis vectors $b_3$ and $b_4$ in this case
   \bea
   b_3&=&\{~|~\tilde{\chi}^{3456}~\tilde \omega_{3456}\},\nonumber\\
 b_4&=&\{~|~\tilde{\chi}^{1256}~\tilde\omega_{1256}  \}\nonumber
 \eea
is incompatible with a 2-shift or a 4-shift $\tilde S_1$ or $\tilde S_2$ because the modular invariance constraints related to $b_3$ and $b_4$ cannot be satisfied at the same time. Hence, the $\left(\begin{array}{c}  0 \\  0 \end{array}\right)$ block can only be combined with $\left(\begin{array}{c}  6 \\  0 \end{array}\right)$, leading to model 3 in Table \ref{tabmodels}.  By symmetry we also get model 2.  Both models belong to equivalence classes of 3 elements.

   Now we look at the possibility of combining the building block $\left(\begin{array}{c} x \\ 0 \end{array}\right)$ to itself or one of the remaining blocks. Models with $\left(\begin{array}{c}  \tilde S_1 \\  \tilde S_2 \end{array}\right)=\left(\begin{array}{c} x \\ 0 \end{array}\right) \rm{and} \left(\begin{array}{c}  S_3 \\  S_4 \end{array}\right)=\left(\begin{array}{c} y \\ 0 \end{array}\right)$ will lead to equivalence classes of 9 elements, since each building block can be rewritten in 3 ways.
There are 9 cases corresponding to $x=2,4,6$ and $y=2,4,6$. Out of these only those for which $(x,y)$ is of the form $(2,2), (4,4), (6,6), (2,4), (4,2)$ are modular invariant.

 For $(x,y)=(2,2)$ the basis vectors take the form
  \bea
 b_1&=&\{ \chi^{3456} ~ y_{12}~\omega_{1...6} |~ \tilde{S}_1\},\nonumber\\
 b_2&=&\{ \chi^{1256} ~  y_{56}~\omega_{12}~~ |~\},\nonumber\\
 b_3&=&\{~~~~~~~~~~~~~~ S_3~|~\tilde{\chi}^{3456}~\tilde y_{12}\tilde \omega_{1...6}\},\nonumber\\
 b_4&=&\{~~~~~~~~~~~~~~~~~~|~\tilde{\chi}^{1256}~\tilde y_{56} \tilde \omega_{12}  \}.\nonumber
 \eea
Compatibility of $\tilde S_1$ with $b_4$ and $S_3$ with $b_2$ forces both shifts to be in the 34 directions, leading to model 5.

Similarly, in the case (4,4) the shifts must be in the 1256 directions leading to model 6,
while the case (6,6) yields model 4.
Case (2,4) is also unique (model 8), since the 2-shift and 4-shift are again restricted to one single possible value, $\tilde S_1 = \tilde y_{34} \tilde \omega_{34}$ and $S_3=y_{1256}\omega_{1256}$. By symmetry we also obtain the case (4,2) called model 7.

Case (6,4), for instance, fails because either $b_1=\{ \chi^{3456} ~ y_{12}~\omega_{1...6} |~ \tilde y_{1...6} \tilde \omega_{1...6}\}$ or $b_2=\{ \chi^{1256} ~  y_{56}\omega_{12} |~\} $ have an odd (complex) intersection in the left part with $b_3=\{ S_3|~\tilde{\chi}^{3456}~\tilde \omega_{3456}\}$ \footnote{$b_2$ requires a  shift in 1256, while $b_1$ requires a shift in 3456.}.
Similar arguments apply for (6,2), (2,6), (4,6).

 If $\left(\begin{array}{c}  S_3 \\  S_4 \end{array}\right)=\left(\begin{array}{c} x \\ 0 \end{array}\right)$ we cannot find two distinct 2-shifts or 4-shifts $\tilde S_1$ and $\tilde S_2$ that are compatible with $b_4=\{~|~\tilde{\chi}^{1256}~\tilde\omega_{1256}  \}$ or $b_4=\{~|~\tilde{\chi}^{1256}~\tilde y_{56}~\tilde\omega_{12}  \}$. So we can only hope to combine this building block with $\left(\begin{array}{c} 6 \\ 2 \end{array}\right)$.
 For $x=2$ we obtain model 11, which is part of a class of 18 elements, because $B_3$ is trivial in this case. By symmetry we also obtain model 9.
Similarly, for $x=4$ we find the models 10 and 12, which are also part of equivalence classes of 18 elements.
No solution is modular invariant when $x=6$.

Coming to the case $\left(\begin{array}{c} 2_x \\ 2_y \end{array}\right)$,  let's look at the option
$\left(\begin{array}{c}  \tilde S_1 \\  \tilde S_2 \end{array}\right)=\left(\begin{array}{c} 2_x \\ 2_y \end{array}\right)$ and $\left(\begin{array}{c}  S_3 \\  S_4 \end{array}\right)=\left(\begin{array}{c} 2_z \\ 2_t \end{array}\right)$, with $x \neq y$ and $z \neq t$. When $\tilde S_1$ is a 2-shift  there is only one option for the left part of $b_1$, $b_{1L}^{ (2)}$, and only one option for the left part of $b_2$ compatible both with a 2-shift in the right part of $b_2$ and with $b_1$. The same option for the left part of $b_1$ is mirrored for the right part of $b_3$ (and similarly $b_2$ is mirrored by $b_4$). There are two independent and modular invariant models.
 One of them has $L=R$ and so it gives rise to a class of 18 elements, model 15, while the other gives rise to a class of 36 elements, model 14.
The remaining choices for the 2-shifts are either already contained in these two classes or do not satisfy modular invariance.

Combining $\left(\begin{array}{c} 2_x \\ 2_y \end{array}\right)$ with $\left(\begin{array}{c} 4_{xy} \\ 4_{xz} \end{array}\right)$ we also find only two independent cases, model 16 and model 17. Both models exhibit the symmetry $L=R$ and are part of equivalence classes of 18 elements.  No solution is obtained when combining $\left(\begin{array}{c} 2_x \\ 2_y \end{array}\right)$ with $\left(\begin{array}{c} 6 \\ 2 \end{array}\right)$.

If all four shifts $\tilde S_{1,2}, ~S_{3,4}$ are 4-shifts we get one model, number  18, from a class of 6 elements, since $L=R$, $B_1=B_3$ and $B_2=B_4$ (hence the only independent transformations are $I, L, B_1, B_2, B_1L, B_2 L$). Next we note that
$\left(\begin{array}{c} 4_{xy} \\ 4_{xz} \end{array}\right)$ is not compatible with $\left(\begin{array}{c} 6 \\ 2 \end{array}\right)$.

Finally combining $\left(\begin{array}{c} 6 \\ 2 \end{array}\right)$ with itself we obtain a class of 36 elements, namely model~\nobreak13.

\section{Appendix: Discrete torsions}
\label{DTsection}
In this appendix we discuss mainly the effect of turning on discrete torsions in model 13. At the end of this section we also present an instance of a chiral model obtained via Gepner map from a Type II model with $h_{11}^{eff}=h_{21}^{eff}$.

As explained in section~\ref{subs:ourapproach} there are six independent discrete torsions in our setup, aside from the untwisted projections.  In the following, we use the labeling
$$ c_{ij}=C\left(\begin{array}{c}  b_i  \\  b_j \end{array}\right)\ , \quad i<j = 1,..,4.$$
Since our models feature only $\Z_2$ twists, the discrete torsions can only take the values $\pm1$.

Because the effect of turning on discrete torsion is very model dependent, we explore in full detail the outcome only for model 13. The choice is motivated by the fact that model 13 contains three NAHE-like twisted sectors that contribute chiral states. Below, we reproduce the basis sets that generate model 13:
\bea
F &=& \{ \psi^\mu\, \chi^{1\ldots 6} \, y^{1\ldots 6} \,
 \omega^{1\ldots 6}  | \, \tilde{y}^{1\ldots 6}\,\tilde{ \omega}^{1\ldots 6}\, \tilde\chi^{1\ldots 6}\, \tilde\psi^{1...10}\, \tilde\phi^{1...16}   \} \ , \nn\\
 S & =& \{\psi^\mu\, \chi^{1\ldots 6}  \}\ , \nn\\
  {E} &=&\{ \tilde\phi^{1...16} \}\ , \nn\\
 b_1&=&\{ \chi^{3456} ~ y_{12}~\omega_{1...6}~ |~ \tilde y_{1...6}\tilde\omega_{1...6}\}\ ,\nonumber\\
 b_2&=&\{ \chi^{1256} ~  y_{34}~\omega_{1...6}~ |~\tilde y_{12}\tilde\omega_{12}\}\ ,\nonumber\\
b_3&=&\{ y_{1...6}~\omega_{1...6}~|~\tilde{\chi}^{3456}~\tilde y_{12}~\tilde \omega_{1...6}\}\ ,\nonumber\\
 b_4&=&\{y_{12}~\omega_{12}~|~\tilde{\chi}^{1256}~\tilde y_{34}~\tilde\omega_{1...6}  \}\ .\nonumber
 \eea
  The untwisted sector is independent of discrete torsions and gives rise to
  \begin{itemize}

\item gauge bosons of $ SO(2)^7 \times SO(4) \times SO(10) \times E^{hidden}_8 $ ;


\item  vector multiplets in the bi-fundamentals
$$(2, 1^4, 2, 1^4),  ~(1, 2, 1^4, 2, 1^3), ~(1,2, 1^5, 4, 1^2) , ~(1^2, 2, 1, 2, 1^5) , ~(1^3, 2, 1^4, 10, 1) , ~(1^6, 2, 4, 1^2)\ .$$
 \end{itemize}
The following twisted  sectors contribute massless states
\begin{itemize}
\item sectors contributing spinorials of $SO(10)$

$\begin{array}{ll}
 \alpha_1=F+b_1+{E}\ ,\hspace{3cm} & \alpha_{123}=F+b_1 + b_2+b_3+{E}\ ,\\
\alpha_3=F+b_3+{E}\ , \hspace{3cm}& \alpha_{134}=F+b_1 + b_3 + b_4 + {E}\ ,\\
\alpha_{24}=F+b_2+ b_4+{E}\ , \hspace{3cm}& \alpha_{1234}= F+b_1+ b_2+ b_3+ b_4+{E}\ .
\end{array}$

\item sectors contributing vectorials of $SO(10)$

$\begin{array}{ll}
\beta_{12}=b_1+ b_2\ ,\hspace{4 cm} & \beta_{23}=b_2 +b_3\ ,\\
\beta_{13}=b_1 +b_3\ ,\hspace{4 cm} & \beta_{124}=b_1+ b_2 +b_4\ ,\\
\beta_{14}=b_1 +b_4\ ,\hspace{4 cm} & \beta_{234}=b_2+ b_3+ b_4\ .
\end{array}$
\end{itemize}

 Table \ref{DT, all+} details the form of the states in each of the twisted sectors, the effect of turning on discrete torsions on the states and, as an example, the spectrum for the case with no discrete torsion. The $SO(2)$ charges in the spinorial and anti-spinorial representations are denoted by $q$ and $\bar q$. The gauge groups $SO(2)_1,~ SO(2)_2 ~\rm {and}~ SO(2)_6$ always appear as $q+\bar q$ in the spinorial representations. This happens because the right moving fermions generating the mentioned gauge groups are always accompanied by a left moving complex fermion and, as a result, the charge of the corresponding Ramond vacuum is not fixed.

 Sectors $\alpha_1$, $\alpha_3$ and $\alpha_{1234}$ always contribute an equal number of $16$'s and $\overline{16}$'s. On the other hand, the chirality of the states arising in sectors $\alpha_{24}$, $\alpha_{123}$ and $\alpha_{134}$ is governed by the same factor $c_{12}c_{14}c_{23}c_{24}c_{34}=\prod_{i < j=1}^{4} c_{ij}/c_{13}$. This occurs because of the following relations
 \bea (F+b_2+b_4+{E})\cap (F+b_1+b_2+b_3+{E})=\{ \psi^\mu ~|~\tilde \psi^{1...10} \}\ ,\nonumber\\
 (F+b_2+b_4+{E})\cap (F+b_1+b_3+b_4+{E})=\{ \psi^\mu ~|~\tilde\psi^{1...10} \}\ ,\label{DTint}
 \eea
that can be interpreted as follows: in sector $\alpha_{24}$ the charge of the fermions $\tilde \psi^{1...10}$ is determined by the projections by $F$, $b_1$, $b_2$, $b_3$ and ${E}$ (or equivalently the projections by $F$, $b_1$, $b_3$, $b_4$ and ${E}$), while in sectors $\alpha_{123}$ and  $\alpha_{134}$ the same charge is controlled by the projections by $F$, $b_2$, $b_4$ and ${E}$. Taking into account the relation between discrete torsion coefficients $ C\left(\begin{array}{c}  a \\  b+c \end{array}\right)=\delta_a~C\left(\begin{array}{c}  a \\  b \end{array}\right) \, C\left(\begin{array}{c}  a \\  c \end{array}\right)$, where $\delta_a$ relates to spin statistics, one can see that the relevant discrete torsion coefficients are indeed $c_{12}$, $c_{14}$, $c_{23}$, $c_{24}$ and $c_{34}$. As a result, the model is chiral for all choices of discrete torsion.

 Other relations similar to (\ref{DTint}) explain the repetitions in the discrete torsion coefficients that govern various gauge group representations in different twisted sectors. It is also interesting to note that the individual values of $c_{12}$ and $c_{24}$ are irrelevant, only their product $c_{12}c_{24}$ matters, as shown in Table  \ref{DT, all+}. This means that varying the discrete torsions leads to only $2^5$ different spectra.
 
 Let us also mention the effect of varying the discrete torsion in a heterotic model obtained via the Gepner map from the Type II model with $(h_{11}^{eff},h_{21}^{eff})=(9,9)$. The sets of the Type II (9,9) model correspond to our model 14 in Table \ref{tabmodels}. The result of the Gepner map can be obtained by adding to model 14 the Wilson line 
 $$G=\{  \tilde \chi^{1...6}  \tilde \psi^{1...10} \},$$
which, as explained in section \ref{Sec:freefermions},
 performs the separation of the compact degrees of freedom and leads to the standard embedding. Via the Gepner map, $G$ is the correspondent of the set $\tilde S$ from Type II. Changing the sign of the discrete torsion $C(1,G)$ with respect to the value of $C(1,S)$ leads from a model with $9 \times 16 ~\rm{and}~ 9 \times \overline{16}$ to a model with $15 \times 16 ~\rm{and}~ 3 \times \overline{16}$. Incidentally, the same effect is obtained when reversing the discrete torsions $C(G,b_3), ~C(G,b_4)$ at the same time from 1 to -1.

{\footnotesize

\begin{landscape}

\begin{longtable}{|c|c|l|c|}

\caption{Discrete torsions in model 13.}
\label{DT, all+}\\

\hline
\textit{Sector}  & \textit{Form of the states} & \textit{~~~~~~~~~~~~Discrete Torsion rules} & \textit{All+}\\
\hline
\endfirsthead

\multicolumn{4}{c}{\bfseries \tablename \ \thetable{} -- continued from previous page}\\

\hline
\textit{Sector}  & \textit{Form of the states} & \textit{~~~~~~~~~~~~Discrete Torsion rules} & \textit{All+}\\
\hline
\endhead

\hline\multicolumn{4}{|r|}{Continued on next page}\\
\hline
\endfoot

\hline
\hline
\endlastfoot

\textbf{ }& $ $ &  $SO(2)_7\times SO(10)$ : $c_{13}c_{14}=1\Rightarrow$  $(q, 16)+(\bar q , \overline{16})$  &   \\
\textbf{$\alpha_1$} & $ (1^4, \ q \ {\rm or}\  \bar q\ , \ q+\bar q\ , \ q \ {\rm or}\  \bar q \ ,\ 1\ ,\ 16 \ {\rm or}\  \overline{16}\ ,\ 1 )$ &  \hspace*{2.65cm} $c_{13}c_{14}=-1 \Rightarrow$  $(q, \overline{16})+(\bar q , 16)$ &  $(1^4, \ q \ , \ q+\bar q\ , \ q  \ ,\ 1\ ,\ 16 \ ,\ 1 ) $ \\
\textbf{} & $(1^4, \ q \ {\rm or}\  \bar q\ , \ q+\bar q\ , \ q \ {\rm or}\  \bar q \ ,\ 1\ ,\ 16 \ {\rm or}\  \overline{16}\ ,\ 1 )$ & $SO(2)_5\times SO(2)_7$ : $c_{14}=1\Rightarrow$  $(q, q)+(\bar q , \bar q)$ &  $(1^4, \ \bar q \ , \ q+\bar q\ , \ \bar q  \ ,\ 1\ ,\ \overline{16} \ ,\ 1 ) $\\
\textbf{} & $$ & \hspace*{2.75cm} $c_{14}=-1\Rightarrow$  $(q, \bar q)+(\bar q , q)$ & \\

\hline
\textbf{} & $$ & $SO(2)_4\times SO(10)$ : $c_{34}=1\Rightarrow$ $(q, 16)+(\bar q , \overline{16})$ &  $(1^2, \ q \ , \ q, \ q  \ ,\ 1^2\, \ 1\ ,\ 16 \ ,\ 1 ) $\\
\textbf{$\alpha_3$} & $(1^2, \ q \ {\rm or}\  \bar q , \ q \ {\rm or}\  \bar q , \ q \ {\rm or}\  \bar q  ,\ 1^2 ,\ 1 ,\ 16 \ {\rm or}\  \overline{16},\ 1 )$ & \hspace*{2.75cm} $c_{34}=-1\Rightarrow$  $(q, \overline{16})+(\bar q , 16)$& $(1^2, \ \bar q \ , \ \bar q, \ \bar q  \ ,\ 1^2\, \ 1\ ,\ \overline{16} \ ,\ 1 ) $\\
& & $SO(2)_3\times SO(2)_5$ :  $c_{34}=1\Rightarrow$ $(q, q)+(\bar q , \bar q)$& $(1^2, \ q \ , \ \bar q, \ q  \ ,\ 1^2\, \ 1\ ,\ \overline{16} \ ,\ 1 ) $\\
\textbf{} & $$ & \hspace*{2.75cm} $c_{34}=-1\Rightarrow$   $(q, \bar q)+(\bar q ,  q)$ & $(1^2, \ \bar q \ , \ q, \ \bar q  \ ,\ 1^2\, \ 1\ ,\ 16 \ ,\ 1 ) $\\
\hline
& & $SO(2)_7$ : $c_{12}c_{13}c_{24}c_{34}=1(-1) \Rightarrow \bar q (q)$ & \\
\textbf{} & $(1, \ q  + \bar q, \ q \ {\rm or}\ \bar q , \ 1^3, \ \ q \ {\rm or}\  \bar q  ,\ 1,\ 16 \ {\rm or}\  \overline{16} ,\ 1 )$ & $SO(2)_3 \times SO(10):$ &  $(1, \ q  + \bar q\ , \ q \ , \ 1^3, \ \   \bar q \ ,\ 1\ ,\ 16 \ ,\ 1 ) $\\
\textbf{$\alpha_{1234}$} & $(1, \ q+ \bar q  , \ \bar q \ {\rm or}\  q , \ 1^3 , \ \bar q \ {\rm or}\  q  ,\ 1 ,\ \overline{16} \ {\rm or}\  16 ,\ 1 )$ & \hspace*{1cm}  $ c_{13}c_{23}c_{34}=1\Rightarrow (q, 16)+(\bar q , \overline{16})$& $(1, \ q+ \bar q \ , \ \bar q \ , \ 1^3\ , \   \bar q \ ,\ 1\ ,\ \overline{16} \ ,\ 1 )$\\
&  & \hspace*{1cm} $c_{13}c_{23}c_{34}=-1\Rightarrow (q, \overline{16})+(\bar q , 16)$ & \\

\hline
\textbf{$\alpha_{24}$} & $(1, \ q  + \bar q, \ 1 ,\ q \ {\rm or}\ \bar q , \ 1, \ q +  \bar q  ,\ 1  ,\ 1 ,\ 16 \ {\rm or}\  \overline{16} ,\ 1 )$ & $SO(2)_4$ : $c_{12}c_{14}c_{24}=1(-1)\Rightarrow \bar q (q)$ & $(1,  \ q  + \bar q , \ 1,\  \bar q , \ 1, \ q +  \bar q  ,\ 1 ,\ 1,\ 16 ,\ 1 )$\\
\textbf{} & $$ &  $SO(10) :  \prod_{i<j=1}^4 c_{ij}/c_{13} =1(-1)\Rightarrow 16(\overline{16})$& \\

\hline
\textbf{$\alpha_{123}$} & $(1^4, \ q \ {\rm or}\ \bar q , \ 1^2, \  2_L +  2_R  ,\ 16 \ {\rm or}\  \overline{16},\ 1 )$ &$SO(2)_5:  \prod_{i<j=1}^4 c_{ij}/c_{23} =1(-1)\Rightarrow q(\bar q)$& $(1^4, \ q \ , \ 1^2, \ \ 2_L +  2_R \ ,\ 16 \ ,\ 1 )$\\
\textbf{} & $$ & $SO(10): \prod_{i<j=1}^4 c_{ij}/c_{13} =1(-1)\Rightarrow 16(\overline{16}) $& \\

\hline
\textbf{} &  & $SO(2)_3 : \prod_{i<j=1}^4 c_{ij}/c_{34} =1(-1)\Rightarrow q(\bar q) $ & \\
\textbf{$\alpha_{134}$} & $2\times(q+\bar{q} ,\ 1, \ q \ {\rm or}\ \bar q , \ 1^3, \  q \ {\rm or}\   \bar q  ,\ 1  ,  16 \ {\rm or}\  \overline{16} , 1 )$ & $SO(2)_7: \prod_{i<j=1}^4 c_{ij}/c_{14}=1(-1)\Rightarrow q(\bar q)$& $  2\times(q+\bar{q}\ ,\ 1, \ q \ , \ 1^3, \ \ q \ ,\ 1 \ , \ 16  ,\ 1 )$ \\
& & $SO(10): \prod_{i<j=1}^4 c_{ij}/c_{13} =1(-1)\Rightarrow 16 (\overline{16})$ & \\

\hline
\hline
\multirow{6}{*}{$\beta_{12}$} & $ (2  , \   1 , \      q \ {\rm or}\ \bar q ,   \  q \ {\rm or}\ \bar q , \    1 , \    1 , \   1 , \      2_L+2_R  , \    1 , \    1  )$ & $SO(2)_3$ : $\prod_{i<j=1}^4 c_{ij}/c_{34}=1(-1) \Rightarrow \bar q (q) $  & $ (2  , \   1 , \  \bar q ,   \  \bar q , \    1 , \    1 , \   1 , \      2_L+2_R  , \    1 , \    1  )$\\
 &  $(1  , \    1 , \      q \ {\rm or}\ \bar q  ,   \  q \ {\rm or}\ \bar q , \     1 , \    1 , \     2   , \       2_L+2_R   , \     1 , \    1  ) $ & $SO(2)_4$ :  $c_{12}c_{14}c_{24}=1(-1)\Rightarrow \bar q (q) $&  $(1  , \    1 , \   \bar q  ,   \   \bar q , \     1 , \    1 , \     2   , \       2_L+2_R   , \     1 , \    1  ) $ \\
 \cline{2-4}
 &  $(1  , \    2 , \     q \ {\rm or}\ \bar q  ,   \  q \ {\rm or}\ \bar q  , \    1 , \    1 , \   1 , \       2_L+2_R   , \    1 , \    1  )$ & $SO(2)_3$ : $\prod_{i<j=1}^4 c_{ij}/c_{34}=1(-1) \Rightarrow q(\bar q)  $&  $(1  , \    2 , \     q  ,   \  q , \    1 , \    1 , \   1 , \       2_L+2_R   , \    1 , \    1  )$\\
& $(1 , \    1 , \     q \ {\rm or}\ \bar q , \  \   q \ {\rm or}\ \bar q  , \    1   , \     2   , \    1  , \      2_L+2_R  , \     1 , \    1 ) $ & $SO(2)_4$ :  $c_{12}c_{14}c_{24}=1(-1)\Rightarrow q(\bar q) $& $(1 , \    1 , \     q , \  \   q   , \    1   , \     2   , \    1  , \      2_L+2_R  , \     1 , \    1 ) $\\
\cline{2-4}
 &  $(1  , \    1 , \      q \ {\rm or}\ \bar q ,   \  q \ {\rm or}\ \bar q , \    2  , \    1 , \    1 , \     2_L+2_R   , \    1 , \    1  )$ & $SO(2)_3$ : $\prod_{i<j=1}^4 c_{ij}/c_{34}=1(-1) \Rightarrow q( \bar q) $ &  $(1  , \    1 , \      q \  ,   \  \bar q , \    2  , \    1 , \    1 , \     2_L+2_R   , \    1 , \    1  )$\\
 & & $SO(2)_4$ :  $c_{12}c_{14}c_{24}=1(-1)\Rightarrow \bar q (q) $ & \\
 \cline{2-4}
 & $(1  , \    1 , \     q \ {\rm or}\ \bar q  ,  \   q \ {\rm or}\ \bar q  , \    1 , \    1 , \    1   , \     2_L+2_R   , \     10   , \   1 )$ & $SO(2)_3$ : $\prod_{i<j=1}^4 c_{ij}/c_{34}=1(-1) \Rightarrow \bar q (q) $& $(1  , \    1 , \   \bar q  ,  \   q   , \    1 , \    1 , \    1   , \     2_L+2_R   , \     10   , \     1  )$ \\
 & & $SO(2)_4$ :  $c_{12}c_{14}c_{24}=1(-1)\Rightarrow q(\bar q) $ & \\

 &  & $SO(2)_3\times SO(2)_4$ : $c_{13}=1\Rightarrow (q, q)+(\bar q ,  \bar q) $ &  \\
& $(2  , \   1 , \      q \ {\rm or}\ \bar q ,    \ q \ {\rm or}\ \bar q , \    1 , \    q+ \bar q ,     \ q \ {\rm or}\ \bar q  , \      1   , \    1 , \    1  ) $ & \hspace*{2.75cm}$c_{13}=-1\Rightarrow (q, \bar q)+(\bar q ,   q)$ & $(2  , \   1 , \      q ,    \ q  , \    1 , \    q+ \bar q ,     \ \bar q  , \      1   , \    1 , \    1  ) $\\
& & $SO(2)_3\times SO(2)_7$ : $c_{14}c_{34}=1\Rightarrow (q, \bar q)+(\bar q ,  q)$ & $(2  , \   1 , \       \bar q ,    \  \bar q , \    1 , \    q+ \bar q ,     \ q   , \      1   , \    1 , \    1  ) $\\
& & \hspace{2.75cm}   $c_{14}c_{34}=-1\Rightarrow (q, q)+(\bar q ,  \bar q)$ & \\

 \cline{2-4}
  \multirow{5}{*}{$\beta_{13}$} & & $SO(2)_3\times SO(2)_4$ : $c_{13}=1\Rightarrow (q, q)+(\bar q ,  \bar q) $ & \\
 &  $(1 ,  2 ,   \  q \ {\rm or}\ \bar q,   \ q \ {\rm or}\ \bar q,  \ 1,  \ q+ \bar q,   \  q \ {\rm or}\ \bar q ,  \   1  ,  \ 1,  \ 1  )  $ & \hspace*{2.75cm}$c_{13}=-1\Rightarrow (q, \bar q)+(\bar q ,   q)$  &  $(1 ,  2 ,   \  q ,   \ q ,  \ 1,  \ q+ \bar q,   \  q  ,  \   1  ,  \ 1,  \ 1  )$\\
 &  $(1  , \   1 , \      q \ {\rm or}\ \bar q ,    \ q \ {\rm or}\ \bar q , \    1 , \    q+ \bar q ,    \ q \ {\rm or}\ \bar q  , \      4   , \    1 , \    1  )$ & $SO(2)_3\times SO(2)_7$ : $c_{14}c_{34}=1\Rightarrow  (q, q)+(\bar q ,  \bar q)$ &  $(1 ,  2 ,   \   \bar q,   \ \bar q,  \ 1,  \ q+ \bar q,   \  \bar q ,  \   1  ,  \ 1,  \ 1  )$\\
 & & \hspace{2.75cm}   $c_{14}c_{34}=-1\Rightarrow(q, \bar q)+(\bar q ,  q)$ & \\

 \cline{2-4}
 &  & $SO(2)_3\times SO(2)_4$ : $c_{13}=1\Rightarrow (q, \bar q)+(\bar q ,   q)$ & \\
& $(1  , \   1 , \      q \ {\rm or}\ \bar q ,     \  q \ {\rm or}\ \bar q , \    2  , \    q+ \bar q , \    \ q \ {\rm or}\ \bar q  , \      1   , \    1 , \    1  ) $ & \hspace*{2.75cm}$c_{13}=-1\Rightarrow (q, q)+(\bar q ,  \bar q) $ & $(1  , \   1 , \      q ,     \   \bar q , \    2  , \    q+ \bar q , \    \ q  , \      1   , \    1 , \    1  ) $\\

& & $SO(2)_3\times SO(2)_7$ : $c_{14}c_{34}=1\Rightarrow (q, q)+(\bar q ,  \bar q)$ & $(1  , \   1 , \       \bar q ,     \  q  , \    2  , \    q+ \bar q , \    \  \bar q  , \      1   , \    1 , \    1  ) $\\
& & \hspace{2.75cm}   $c_{14}c_{34}=-1\Rightarrow (q, \bar q)+(\bar q ,  q)$& \\

\cline{2-4}
 &  & $SO(2)_3\times SO(2)_4$ : $c_{13}=1\Rightarrow (q, \bar q)+(\bar q ,   q)$ & \\
&   $  (1  , \   1 , \      q \ {\rm or}\ \bar q ,    \ q \ {\rm or}\ \bar q , \    1 , \    q+ \bar q ,    \ q \ {\rm or}\ \bar q  , \      1   , \    10 , \    1  )$ & \hspace{2.75cm}$c_{13}=-1\Rightarrow (q, q)+(\bar q ,  \bar q)$ &   $  (1  , \   1 , \      q  ,    \  \bar q , \    1 , \    q+ \bar q ,    \ \bar q  , \      1   , \    10 , \    1  )$\\
& & $SO(2)_3\times SO(2)_7$ : $c_{14}c_{34}=1\Rightarrow (q, \bar q)+(\bar q ,  q)$ &   $  (1  , \   1 , \   \bar q ,    \ \bar q , \    1 , \    q+ \bar q ,    \ q , \      1   , \    10 , \    1  )$\\
& & \hspace{2.75cm}   $c_{14}c_{34}=-1\Rightarrow (q, q)+(\bar q ,  \bar q)$ & \\

\hline

 &  & $SO(2)_4: c_{12}c_{14}c_{24}=1(-1)\Rightarrow q( \bar q)$ & \\
&   $2\times     (q+\bar q , \ 2,  \ 1,  \  q \ {\rm or}\ \bar q,   \ q \ {\rm or}\ \bar q,  \ 1,  \ q \ {\rm or} \ \bar q,   \   1  ,  \ 1,  \ 1  )$ & $SO(2)_5: \prod_{i<j=1}^4 c_{ij}/c_{23}=1(-1)\Rightarrow q( \bar q)$&   $2\times     (q+\bar q , \ 2,  \ 1,  \  q,   \ q ,  \ 1,  \ q,   \   1  ,  \ 1,  \ 1  )$\\
& & $SO(2)_7: c_{12}c_{13}c_{24}c_{34}=1(-1)\Rightarrow q( \bar q)$ & \\

\cline{2-4}
&  & $SO(2)_4: c_{12}c_{14}c_{24}=1(-1)\Rightarrow \bar q(q)$ & \\
&   $2\times    (q+\bar q , \ 1,  \ 2 ,  \  q \ {\rm or}\ \bar q,   \ q \ {\rm or}\ \bar q,  \ 1,  \ q \ {\rm or} \ \bar q,   \   1  ,  \ 1,  \ 1  )$ & $SO(2)_5: \prod_{i<j=1}^4 c_{ij}/c_{23}=1(-1)\Rightarrow q( \bar q)$  &   $2\times    (q+\bar q , \ 1,  \ 2 ,  \   \bar q,   \ q,  \ 1,  \ q ,   \   1  ,  \ 1,  \ 1  )$\\
& & $SO(2)_7: c_{12}c_{13}c_{24}c_{34}=1(-1)\Rightarrow q( \bar q)$ & \\

\cline{2-4}
$\beta_{14}$   &  & $SO(2)_4: c_{12}c_{14}c_{24}=1(-1)\Rightarrow q( \bar q)$ & \\
&   $2\times    (q+\bar q , \ 1 ,  \ 1,  \  q \ {\rm or}\ \bar q,   \ q \ {\rm or}\ \bar q,  \ 2,  \ q \ {\rm or} \ \bar q,   \   1  ,  \ 1,  \ 1  )$ & $SO(2)_5: \prod_{i<j=1}^4 c_{ij}/c_{23}=1(-1)\Rightarrow q( \bar q)$ &   $2\times    (q+\bar q , \ 1 ,  \ 1,  \  q ,   \   q,  \ 2,  \ \bar q ,   \   1  ,  \ 1,  \ 1  )$\\
& & $SO(2)_7: c_{12}c_{13}c_{24}c_{34}=1(-1)\Rightarrow \bar q(q)$ & \\

\cline{2-4}
 &  & $SO(2)_4: c_{12}c_{14}c_{24}=1(-1)\Rightarrow \bar q(q)$ & \\
&   $2\times     (q+\bar q , \ 1 ,  \ 1,  \  q \ {\rm or}\ \bar q,   \ q \ {\rm or}\ \bar q,  \ 1,  \ q \ {\rm or} \ \bar q,   \   4  ,  \ 1,  \ 1  )$ & $SO(2)_5: \prod_{i<j=1}^4 c_{ij}/c_{23}=1(-1)\Rightarrow \bar q( q) $&   $2\times     (q+\bar q , \ 1 ,  \ 1,  \  \bar q,   \  \bar q,  \ 1,  \  \bar q,   \   4  ,  \ 1,  \ 1  )$ \\
& & $SO(2)_7: c_{12}c_{13}c_{24}c_{34}=1(-1)\Rightarrow \bar q(q)$ & \\

\cline{2-4}
&  & $SO(2)_4: c_{12}c_{14}c_{24}=1(-1)\Rightarrow q( \bar q)$ &  \\
&   $2\times    (q+\bar q , \ 1,  \ 1,  \  q \ {\rm or}\ \bar q,  \ q \ {\rm or}\ \bar q,  \ 1,  \ q \ {\rm or} \ \bar q,   \   1  ,  \ 10,  \ 1  )$ & $SO(2)_5: \prod_{i<j=1}^4 c_{ij}/c_{23}=1(-1)\Rightarrow \bar q(q)$ &   $2\times    (q+\bar q , \ 1,  \ 1,  \  q,  \ \bar q ,  \ 1,   \  q,   \   1  ,  \ 10,  \ 1  )$\\
& & $SO(2)_7: c_{12}c_{13}c_{24}c_{34}=1(-1)\Rightarrow q( \bar q)$ & \\

\hline
 &  & $SO(2)_7:\prod_{i<j=1}^4 c_{ij}/c_{14}=1(-1)\Rightarrow \bar q(q)$ &  \\
& $(2,  \ q+\bar q ,  \ 1,  \  q \ {\rm or}\ \bar q,   \ q \ {\rm or}\ \bar q,  \ 1,  \ q \ {\rm or} \ \bar q,   \   1  ,  \ 1,  \ 1  )  $ & $SO(2)_4\times SO(2)_5: c_{13}c_{23}c_{34}=1\Rightarrow (q, q)+(\bar q ,  \bar q) $ &  $(2,  \ q+\bar q ,  \ 1,  \  q ,   \ q ,  \ 1,  \  \bar q,   \   1  ,  \ 1,  \ 1  )  $\\
& & \hspace{2.55cm}   $c_{13}c_{23}c_{34}=-1\Rightarrow (q, \bar q)+(\bar q ,  q)$ & $(2,  \ q+\bar q ,  \ 1,  \   \bar q,   \  \bar q,  \ 1,  \  \bar q,   \   1  ,  \ 1,  \ 1  )  $\\

\cline{2-4}
 $\beta_{124}$& $ (1,  \ q+\bar q ,  \ 1,  \  q \ {\rm or}\ \bar q,   \ q \ {\rm or}\ \bar q,  \  2,  \ q \ {\rm or} \ \bar q,   \   1  ,  \ 1,  \ 1  )$ & $SO(2)_7:\prod_{i<j=1}^4 c_{ij}/c_{14}=1(-1)\Rightarrow q(\bar q)$ & $(1,  \ q+\bar q ,  \ 1,  \  q ,   \  q,  \  2,  \   q,   \   1  ,  \ 1,  \ 1  )$ \\
& & $SO(2)_4\times SO(2)_5: c_{13}c_{23}c_{34}=1\Rightarrow (q, q)+(\bar q ,  \bar q) $ & $(1,  \ q+\bar q ,  \ 1,  \  \bar q,   \ \bar q,  \  2,  \ q ,   \   1  ,  \ 1,  \ 1  )$ \\
&  $(1,  \ q+\bar q ,  \ 1,  \  q \ {\rm or}\ \bar q,   \ q \ {\rm or}\ \bar q,  \ 1,  \ q \ {\rm or} \ \bar q,   \   4  ,  \ 1,  \ 1  )  $ & \hspace{2.55cm}   $c_{13}c_{23}c_{34}=-1\Rightarrow (q, \bar q)+(\bar q ,  q)$ &  $(1,  \ q+\bar q ,  \ 1,  \  q ,   \  q,  \ 1,  \  q,   \   4  ,  \ 1,  \ 1  )  $\\
& & &  $(1,  \ q+\bar q ,  \ 1,  \  \bar q,   \ \bar q ,  \ 1,  \ q,   \   4  ,  \ 1,  \ 1  )  $\\

\cline{2-4}
 & $(1,  \ q+\bar q ,  \  2,  \  q \ {\rm or}\ \bar q,  \ q \ {\rm or}\ \bar q,  \ 1,  \ q \ {\rm or} \ \bar q,   \   1  ,  \ 1,  \ 1  )$ & $SO(2)_7:\prod_{i<j=1}^4 c_{ij}/c_{14}=1(-1)\Rightarrow \bar q(q)$ & $(1,  \ q+\bar q ,  \  2,  \  q ,  \ \bar q ,  \ 1,  \ \bar q,   \   1  ,  \ 1,  \ 1  ) $\\
&  & $SO(2)_4\times SO(2)_5: c_{13}c_{23}c_{34}=1\Rightarrow (q, \bar q)+(\bar q ,   q) $ & $ (1,  \ q+\bar q ,  \  2,  \   \bar q,  \  q,  \ 1,  \ \bar q ,   \   1  ,  \ 1,  \ 1  )$\\
& $(1,  \ q+\bar q ,  \ 1,  \  q \ {\rm or}\ \bar q,  \ q \ {\rm or}\ \bar q,  \ 1,  \ q \ {\rm or} \ \bar q,   \   1  ,  \ 10,  \ 1  )  $ & \hspace{2.55cm}   $c_{13}c_{23}c_{34}=-1\Rightarrow (q,  q)+(\bar q ,  \bar q)$ & $ (1,  \ q+\bar q ,  \ 1,  \  q ,  \ \bar q,  \ 1,  \ \bar q,   \   1  ,  \ 10,  \ 1  )  $\\
& & & $ (1,  \ q+\bar q ,  \ 1,  \  \bar q,  \ q ,  \ 1,  \ \bar q ,   \   1  ,  \ 10,  \ 1  )  $\\

\hline
& $(2,  \ q+\bar q ,   \ q \ {\rm or}\ \bar q,  \ 1 , \  q \ {\rm or}\ \bar q,    \ q + \bar q,  \ 1,   \   1  ,  \ 1,  \ 1  )$ & $SO(2)_3: \prod_{i<j=1}^4 c_{ij}/c_{34}=1(-1)\Rightarrow \bar q(q)$ & $(2,  \ q+\bar q ,   \ \bar q,  \ 1 , \  q ,    \ q + \bar q,  \ 1,   \   1  ,  \ 1,  \ 1  )$ \\
&   $(1,  \ q+\bar q ,    \ q \ {\rm or}\ \bar q,  \ 1 ,  \ q \ {\rm or}\ \bar q,     \ q + \bar q,  \ 2,   \   1  ,  \ 1,  \ 1  )$ & $SO(2)_5: \prod_{i<j=1}^4 c_{ij}/c_{23}=1(-1)\Rightarrow q( \bar q)$&   $(1,  \ q+\bar q ,    \  \bar q,  \ 1 ,  \ q ,     \ q + \bar q,  \ 2,   \   1  ,  \ 1,  \ 1  )$ \\

\cline{2-4}
 $\beta_{234}$ & $(1,  \ q+\bar q ,    \ q \ {\rm or}\ \bar q,  \  2 ,  \ q \ {\rm or}\ \bar q,     \ q + \bar q,  \ 1,   \   1  ,  \ 1,  \ 1  )$ & $SO(2)_3: \prod_{i<j=1}^4 c_{ij}/c_{34}=1(-1)\Rightarrow q(\bar q)$ & $(1,  \ q+\bar q ,    \ q,  \  2 ,  \ q,     \ q + \bar q,  \ 1,   \   1  ,  \ 1,  \ 1  )$\\
&    & $SO(2)_5: \prod_{i<j=1}^4 c_{ij}/c_{23}=1(-1)\Rightarrow q( \bar q)$ & \\

\cline{2-4}
& $(1,  \ q+\bar q ,    \ q \ {\rm or}\ \bar q,  \ 1 ,  \ q \ {\rm or}\ \bar q,     \ q + \bar q,  \ 1,   \   4  ,  \ 1,  \ 1  )$ & $SO(2)_3: \prod_{i<j=1}^4 c_{ij}/c_{34}=1(-1)\Rightarrow q(\bar q)$ & $(1,  \ q+\bar q ,    \ q ,  \ 1 ,  \ \bar q,     \ q + \bar q,  \ 1,   \   4  ,  \ 1,  \ 1  )$ \\
&    & $SO(2)_5: \prod_{i<j=1}^4 c_{ij}/c_{23}=1(-1)\Rightarrow  \bar q(q)$ & \\

\cline{2-4}
& $(1,  \ q+\bar q ,    \ q \ {\rm or}\ \bar q,  \ 1 ,  \ q \ {\rm or}\ \bar q,     \ q + \bar q,  \ 1,   \   1  ,  \ 10,  \ 1  )$ & $SO(2)_3: \prod_{i<j=1}^4 c_{ij}/c_{34}=1(-1)\Rightarrow \bar q(q)$ & $(1,  \ q+\bar q ,    \  \bar q,  \ 1 ,  \  \bar q,     \ q + \bar q,  \ 1,   \   1  ,  \ 10,  \ 1  )$\\
&    & $SO(2)_5: \prod_{i<j=1}^4 c_{ij}/c_{23}=1(-1)\Rightarrow  \bar q(q)$ & \\

\hline
 \hspace{0.25cm} $\beta_{23}$ & $c_{23}=1 : (2, \ 1^4,  \ q+\bar q ,    \ q \ {\rm or}\ \bar q, \ 2_L+2_R,  \ 1^2   )$ & $SO(2)_7: c_{12}c_{13}c_{24}c_{34}=1(-1)\Rightarrow \bar q(q)$\hspace{1.75cm} &  $(2, \ 1, \ 1, \ 1, \ 1,  \ q+\bar q ,    \  \bar q, \ 2_L+2_R,  \ 1^2   )$\\
\cline{2-4}
    &  $c_{23}=1 : (1, \ 2, \ 1^3,  \ q+\bar q ,    \ q \ {\rm or}\ \bar q, \ 2_L+2_R,  \ 1^2   )$ & $SO(2)_7: c_{12}c_{13}c_{24}c_{34}=1(-1)\Rightarrow q(\bar q)$ & $(1, \ 2, \ 1, \ 1, \ 1,  \ q+\bar q ,    \ q, \ 2_L+2_R,  \ 1^2  )$\\
\cline{2-4}
 & $c_{23}=-1 : (1^2 , \ 2,  \ 1^2,  \ q+\bar q ,    \ q \ {\rm or}\ \bar q, \ 2_L+2_R,  \ 1^2   )~ \,$ & $SO(2)_7: c_{12}c_{13}c_{24}c_{34}=1(-1)\Rightarrow q(\bar q)$ & \\
& $ (1^5,  \ q+\bar q ,    \ q \ {\rm or}\ \bar q, \ 2_L+2_R,  \ 10,   \   1   )$ & & \\
\cline{2-4}
  &   $c_{23}=-1 : (1^3, \ 2, \ 1,  \ q+\bar q ,    \ q \ {\rm or}\ \bar q, \ 2_L+2_R,  \ 1^2   )$ & $SO(2)_7: c_{12}c_{13}c_{24}c_{34}=1(-1)\Rightarrow \bar q(q)$ & \\
& $ (1^4, \ 2,  \ q+\bar q ,    \ q \ {\rm or}\ \bar q, \ 2_L+2_R,  \ 1^2   )$ & &  \\

\end{longtable}
\end{landscape}

}


\section{Appendix: $\Z_2 \times \Z_2$  characters}
\label{characters}

In this appendix we report the characters $\tau$ and the amplitudes $\rho$ used for the $\Z_2 \times \Z_2$ twists in the space-time directions. The $\tau$ characters are defined as follows \cite{MBthesis}:
\begin{eqnarray}
\tau_{00} &=& VOOO + OVVV - SCCS - CSSC \ \sim V-S-C \ ;\nonumber \\
\tau_{01} &=& VOVV + OVOO - SCSC - CSCS \ \sim 2O-S-C \ ;\nonumber \\
\tau_{02} &=& VVOV + OOVO - SSCC - CCSS \ \sim 2O-S-C  \ ;\nonumber \\
\tau_{03} &=& VVVO + OOOV - SSSS - CCCC \ \sim 2O-S-C  \ ;\nonumber \\
\tau_{10} &=& OOCS + VVSC - SSOO - CCVV  \ \sim O-S \ ;\nonumber \\
\tau_{11} &=& OOSC + VVCS - SSVV - CCOO \ \sim O-C \ ;\nonumber \\
\tau_{12} &=& OVCC + VOSS - SCOV - CSVO  \ ;\nonumber \\
\tau_{13} &=& OVSS + VOCC - SCVO - CSOV \ ;\nonumber \\
\tau_{20} &=& OCOS + VSVC - SOSO - CVCV  \ \sim O-S \ ;\nonumber \\
\tau_{21} &=& OCVC + VSOS - SOCV - CVSO \ ;\nonumber \\
\tau_{22} &=& OSOC + VCVS - SVSV - COCO \ \sim O-C \ ; \nonumber \\
\tau_{23} &=& OSVS + VCOC - SVCO - COSV \ ;\nonumber \\
\tau_{30} &=& OSSO + VCCV - SVVC - COOS  \ \sim O-C \ ;\nonumber \\
\tau_{31} &=& OSCV + VCSO - SVOS - COVC \ ;\nonumber \\
\tau_{32} &=& OCSV + VSCO - SOVS - CVOC \ ;\nonumber \\
\tau_{33} &=& OCCO + VSSV - SOOC - CVVS \ \sim O-S \ ,
\end{eqnarray}
where we have also indicated the space-time (potential) massless contributions in terms of the transverse Lorentz $SO(2)$ representations in four dimensions.  The corresponding amplitudes are the combinations respecting the $\Z_2 \times \Z_2$ orbifold group structure. They are given by
\bea
\rho_{\alpha 0} &=&  \tau_{\alpha 0}+\tau_{\alpha 1}+\tau_{\alpha 2}+\tau_{\alpha 3} \ ;\nn\\
\rho_{\alpha 1} &=&  \tau_{\alpha 0}+\tau_{\alpha 1}-\tau_{\alpha 2}-\tau_{\alpha 3} \ ;\nn\\
\rho_{\alpha 2} &=&  \tau_{\alpha 0}-\tau_{\alpha 1}+\tau_{\alpha 2}-\tau_{\alpha 3} \ ;\nn\\
\rho_{\alpha 3} &=&  \tau_{\alpha 0}-\tau_{\alpha 1}-\tau_{\alpha 2}+\tau_{\alpha 3} \ .\nn
\eea

In the partition function they must be supported by the internal amplitudes, that can be arranged in terms of the
characters of the corresponding $so(2n)$ affine algebras.

\section{Appendix: The algorithm}
We give here a brief description of the algorithm that was created to scan our $2^{20}$ models and used Wolfram Mathematica $8.0$ as software environment.  The algorithm consists of three modules. The first one generates all the possible fermion configurations describing the four sets of twists/shifts and selects only the ones consistent with worldsheet supersymmetry and modular invariance constraints.  The second module constructs the modular invariant amplitude for a given consistent model.  The third one extracts the spectrum in terms of (super)characters.

\begin{itemize}
\item \textbf{Module 1}. The algorithm generates all the possible quartets of sets related to the fermionized degrees of freedom of the heterotic string as described in Section \ref{Sec:freefermions}. In their raw form the sets consist in arrays of 1 and -1 (-1 for twisted fermions). Then the consistency conditions (\ref{consistency}, \ref{condsusy}) filter only those sets compatible with world-sheet supersymmetry and modular invariance. For each consistent model, the four sets determine $2^4=16$ classes of fermions for left and right modes separately, according to the fact that a single fermion can be twisted along the ``space'' and/or ``time'' direction of the world-sheet. A single model - namely a set of four specific fermion sets - can be also identified with this ordered arrays of integers that correspond precisely to the thetas' exponents in the untwisted amplitude. Two different lists of 4 sets can produce the same array of integers, and in this case these models are considered completely equivalent. On the other hand, two arrays with the same integers in different orders, produce in general the same untwisted amplitude, but different twisted sectors, and, therefore, are not equivalent.

\item \textbf{Module 2}. Once the correspondence between models and integer arrays has been settled, the algorithm works directly with the arrays. The torus partition function, in terms of Jacobi thetas, is splitted as dictated by the integers and projected to produce the untwisted amplitude. Moreover, each power of theta comes with one of the $16$ class label that uniquely identifies the action of the four basis sets. Starting from the initial torus, the full modular invariant Partition Function consists of $256$ amplitudes, due to the $16$-dimensional orbifold group acting on the $16$ ($1$ untwisted and $15$ twisted) sectors. The amplitudes are of two types: the ones reached by a modular transformation ($T$ or $S$) from another amplitude, and the "disconnected" ones.
    As stated in Section \ref{subs:ourapproach},  there are $15$ "orbits" of $6$ elements related by $T$ and $S$ transformations and an untwisted orbit of $46$ elements.

\item \textbf{Module 3}. Each orbit has a phase (1 or -1) hidden in the definition of tau's. To fix the $15$ orbit signs (the untwisted orbit phase is by definition fixed to $1$) we scan the spectra and keep the consistent ones.  They can be recognized, once the Jacobi thetas of the internal part are expanded in q powers and written in terms of characters, by requiring the coefficient of the characters to be integer and positive. Once the consistent signs configuration is found, the spectrum is finally printed out for each sector, separately.
    Of course, not all orbit phases are constrained.  Six of them are independent and correspond to the $2^6 = 64$ discrete torsion variations of the same model. These are exhaustively explored.

\end{itemize}


\end{document}